\documentclass[ aip,jcp,reprint,amsmath,amssymb,preprintnumbers]{revtex4-2}

\usepackage[utf8]{inputenc}
\DeclareUnicodeCharacter{2009}{\,}  

\usepackage{xcolor}
\usepackage{graphicx}
\usepackage{hyperref}
\usepackage{algpseudocode}
\newcounter{myalgorithm}
\renewcommand{\themyalgorithm}{\arabic{myalgorithm}}
\usepackage{amsmath}
\usepackage{amssymb}
\usepackage[capitalise]{cleveref}
\usepackage{booktabs}
\usepackage{multirow}
\usepackage{changepage}
\begin{document}
\preprint{LA-UR-26-29564}

\title{Shadow Molecular Dynamics for General Nonlinear Flexible-Charge Models}

\author{Jos Thomas}
\affiliation{Theoretical Division, Los Alamos National Laboratory,
Los Alamos, New Mexico 87545 USA}
\affiliation{School of Computing, DePaul University, Chicago, Illinois 60604, USA}
\author{Anthony P. Baldo}
\affiliation{Theoretical Division, Los Alamos National Laboratory,
Los Alamos, New Mexico 87545 USA}
\author{Rae A. Corrigan Grove}
\affiliation{Theoretical Division, Los Alamos National Laboratory,
Los Alamos, New Mexico 87545 USA}
\author{Mehmet Cagri Kaymak}
\affiliation{Theoretical Division, Los Alamos National Laboratory,
Los Alamos, New Mexico 87545 USA}
\author{Joshua Finkelstein}
\affiliation{Theoretical Division, Los Alamos National Laboratory,
Los Alamos, New Mexico 87545 USA}
\author{James Goff}
\affiliation{Center for Computing Research, Sandia National Laboratories,
Albuquerque, New Mexico 87185 USA}
\author{Maksim Kulichenko}
\affiliation{Theoretical Division, Los Alamos National Laboratory,
Los Alamos, New Mexico 87545 USA}
\author{Sergei Tretiak}
\affiliation{Theoretical Division, Los Alamos National Laboratory,
Los Alamos, New Mexico 87545 USA}
\author{Nicholas Lubbers}
\affiliation{Computing and Artificial Intelligence Division, Los Alamos National Laboratory, Los Alamos, New Mexico 87545 USA}
\author{Vishikh Athavale}
\affiliation{Theoretical Division, Los Alamos National Laboratory,
Los Alamos, New Mexico 87545 USA}
\author{Christian F. A. Negre}
\affiliation{Theoretical Division, Los Alamos National Laboratory,
Los Alamos, New Mexico 87545 USA}
\author{Michael Wall}
\affiliation{Computing and Artificial Intelligence Division, Los Alamos National Laboratory, Los Alamos, New Mexico 87545 USA}
\author{Anders M. N. Niklasson}
\email{amn@lanl.gov}
\affiliation{Theoretical Division, Los Alamos National Laboratory,
Los Alamos, New Mexico 87545 USA}

\date{\today}

\begin{abstract}
Machine-learned interatomic potentials have greatly expanded the reach of molecular dynamics simulations, but complex chemical environments often require explicit treatment of long-range electronic response, including charge transfer and polarization. Physics-based flexible-charge models capture these effects but require global nonlinear charge equilibration at every time step to obtain Born-Oppenheimer energies and forces. This process is computationally expensive and can lead to  instabilities unless tightly converged. Here, we introduce a shadow molecular dynamics framework for flexible-charge models with arbitrary local nonlinearities and long-range multipolar electrostatics. We replace the exact Born-Oppenheimer potential with a nearby shadow potential obtained by exactly equilibrating a partially linearized electronic energy functional. This provides conservative forces and reduces the global nonlinear equilibration to a quasi-diagonal linear problem that can be solved directly at low cost. Auxiliary dynamical electronic degrees of freedom, propagated through an extended-Lagrangian dynamics, capture the long-range electronic response and define the reference state for the linearization of the energy functional. GPU-accelerated simulations with over 100,000 atoms demonstrate that the shadow potential closely follows the tightly converged Born-Oppenheimer surface, while maintaining long-term stability at a computational cost approaching that of a fixed-charge model. These advances establish an efficient and broadly applicable framework for machine-learned interatomic potentials that explicitly incorporate long-range electrostatic interactions and environment-dependent electronic response.
\end{abstract}

\maketitle

\section{Introduction}

Atomistic molecular dynamics (MD) is a fundamental tool for investigating microscopic processes in chemistry, materials science, and molecular biology. \cite{RCar85, MAllen90, PCarloni02, DFrenkel02, JTse02, AVoter02, RIftimie05, marx_hutter_2009, MTuckerman10, MKarplus14}  Machine-learned interatomic potentials (MLIPs) are rapidly transforming MD simulations by combining near-first-principles accuracy with a computational efficiency approaching that of classical force fields. \cite{MRupp12,Anatole15,APThompson15,JBehler16,NSchutt17,JSSmith19,Drautz19,Tkatchenko20,UMA25,Loew2025,Mazitov2025,Xu2025,Wen2025,Kim2026,Lysogorskiy2026} Despite these advances, accurately capturing nonlocal electronic effects, including long-range electrostatics, charge transfer, and polarization, remains a major challenge for modern MLIPs. \cite{Faraji2017,KYao18,TWKo20,MCeriotti19,OUnke19,DPMetcalf21,Shao_2022,Weinan_2022,Staacke2022,AGao22,RahmatizadKhajehpasha2022,TWKo23,Isborn_2023,DAnstine23,JGoff23,DKim2024,CZhang24,Shaidu2024,Li2025,Cheng2025,Kim2025,Thomas_2025,Fuchs2025,Ghasemi2015,Rinaldi2025,Vondrak2025,JYajie25,DAnstine2025,Zhang2026,Vondrak2026}
These effects are essential to describe a broad range of systems, such as electrochemical interfaces, ionic materials, heterogeneous catalysts, defective molecular crystals, and biological environments. \cite{Catlow1983,Grozovski2006,Norskov2009,French2010}
In such systems, the electronic charge distributions are governed by a global electronegativity equilibration involving long-range Coulomb interactions, rather than solely by the local atomic environments. 
Flexible-charge models account for these effects by introducing physics-based electronic degrees of freedom determined through a global charge equilibration \cite{FJVesely77,Gasteiger80,WJMortier86,MSprik88,MSprik90,AKRappe91,DVanBelle92,WSRick94,ACTvanDuin01,TAHalgren01,GLamoureaux03,GAKaminsky04,RNistor06,PEMLopes09,PCieplak09,TVerstraelen13,SNaserifar17,JZhifeng19,YLi19,Lemkul2016,senftle2016reaxff,Jensen2023} based on Sanderson's electronegativity equalization principle. \cite{Sanderson1951}
Combined with machine-learned representations of their local electronic energy contributions, these models provide a physically motivated foundation for constructing MLIPs with long-range electronic response.

The purpose of this article, however, is not to introduce another class of MLIPs optimized for specific molecular systems using new descriptors or neural-network architectures. Instead, our goal is to develop a general theoretical and computational framework for efficient and stable MD simulations with nonlinear flexible-charge models. The framework is independent of specific machine-learning architectures and parameterizations and is applicable to a broad range of physics-based and machine-learned interatomic potentials that incorporate long-range electronic response. Our focus is on the underlying MD methodology rather than on the construction or optimization of the interatomic potentials.

In MD simulations the main challenge to include long-range electronic response is computational.
For each nuclear configuration, the electronic degrees of freedom are determined by minimizing a nonlinear electronic energy functional under a global charge constraint. 
This charge-equilibration problem defines the Born-Oppenheimer potential-energy surface and generally requires iterative nonlinear optimization at every MD time step.
Incomplete convergence may produce nonconservative forces, systematic energy drift, and loss of long-term numerical stability.
To overcome these shortcomings, conventional methods often try to accelerate and improve convergence through extrapolated initial guesses or more efficient nonlinear solvers. In this work we follow a different approach. Instead, the nonlinear, global charge-equilibration problem is replaced by an exactly solvable linear system.

The key conceptual idea is to replace the exact Born-Oppenheimer potential-energy surface with a nearby shadow potential obtained by exactly equilibrating a partially linearized electronic energy functional. \cite{ANiklasson21,ANiklasson21b} This shadow potential provides exact conservative forces by replacing replaces the iterative nonlinear charge-equilibration problem with a structured (quasi-diagonal) linear system that can be solved directly at low computational cost.
This kind of backward error analysis \cite{Wilkinson1963,SReich99,golub2013matrix} is based on the same concept as is used in geometric or symplectic integration schemes motivated by the existence of an underlying shadow Hamiltonian. \cite{HYoshida90,CGrebogi90,SToxvaerd94,BenettinGiorgilli1994,HairerLubichWanner2006,SkeelHardy2001,BJLeimkuhler04} However, the shadow approximation introduced here differs fundamentally from these shadow Hamiltonians, which arise from the time discretization of the equations of motion. Instead, our approximation is constructed at the level of the nonlinear electronic energy functional and is therefore independent of the choice of numerical integration method.

To maintain the accuracy of the shadow approximation during the MD simulations, the partial linearization of the electronic energy functional is performed around an approximate electronic ground state that evolves dynamically together with the nuclei. This is formulated within an extended-Lagrangian framework. \cite{HCAndersen80,ANiklasson08,ANiklasson21b} The extended electronic dynamical variables continuously update the shadow electronic energy functional and need only to remain sufficiently close to the exact ground state for the shadow potential to accurately follow the regular 'exact' reference Born-Oppenheimer potential surface. The computational effort in this shadow MD framework is therefore shifted from repeated tightly converged nonlinear optimizations to the approximate ground-state propagation of extended dynamical variables together with the exact solution of an inexpensive linear charge-equilibration problem.

The present formulation builds on previous developments in shadow MD and extended-Lagrangian Born-Oppenheimer MD for charge-equilibration models, \cite{ANiklasson08,KNomura15,AAlbaugh15,AAlbaugh17,AAlbaugh18,ANiklasson21,ANiklasson21b,JGoff23,RStanton25,Li2025,RGrove26} but introduces some important improvements. In particular, the formulation enables the use of off-the-shelf Krylov-subspace solvers and extends the theory to more general functional forms, including higher-order nonlinear functionals and electrostatic multipole interactions. The modified formulation also improves the computational efficiency and numerical stability under approximations compared to recent shadow MD implementations. \cite{Kaymak2025GPU,Li2025,JGoff23,RGrove26} Furthermore, we demonstrate how the shadow MD can be combined with Langevin dynamics for canonical (NVT) simulations. Many important properties, such as the scaling of the error in the charges, forces and shadow potential as a function of the integration time step,  will remain the same as for previous shadow MD formulations and will not be analyzed here. 

The proposed shadow MD framework is sufficiently general to include established second-order charge-equilibration (QEq) models such as MLIPs based on latent charges and their long-range Coulomb interactions, multipole interaction models, as well as models with higher-order nonlinear charge dependencies. We will use examples based on both conventional second-order and higher-order charge-equilibration models to demonstrate the theory. More general nonlinear energy functionals or higher-order multipolar interactions, as discussed in Ref.\ \cite{RGrove26}, will not be considered in the numerical demonstrations but are expected to exhibit similar behavior.

The article is organized as follows. Section \ref{sec:flexible_charge_model} introduces the Born-Oppenheimer MD formulation for general nonlinear flexible-charge models. In Section \ref{sec:ShadowMD} we develop the shadow MD framework, which is the core part of the article, and in Section \ref{sec:DifferentModels} we present some particular classes of models that can be treated within this framework. Section \ref{sec:Examples} provides numerical demonstrations, including for a fourth-order nonlinear model, but with a particular emphasis on the commonly used second-order QEq model. Section \ref{sec:Perspective} discusses shadow MD from a backward-error perspective and its relation to shadow Hamiltonians. Finally, in Section \ref{sec:Summary} we summarize the main results and present our conclusions.                                                  
\section{Molecular Dynamics for Flexible-Charge Equilibration Models}
\label{sec:flexible_charge_model}

Flexible-charge models determine the equilibrated electronic charge distribution and energy by minimizing a charge-dependent energy functional subject to global charge conservation, or, equivalently, through a global electronegativity equilibration that includes long-range, all-to-all Coulomb interactions. \cite{Sanderson1951} In contrast to fixed-charge models, the charge distribution must therefore be updated for each new atomic configuration during an MD simulation. Flexible-charge models can take a variety of forms. \cite{MSprik88,AKRappe91,ACTvanDuin01,RNistor06,JChen08,MZhang2009,PEMLopes09,TVerstraelen13,Poier2019,Jensen2023,Froitzheim2025} Here, we adopt a fairly general formulation motivated by density functional theory (DFT), \cite{HohenbergKohn64,KohnSham65,Parr89,RMDreizler90,WKohn99} which provides a first-principles theoretical foundation for a broad class of flexible-charge models. \cite{DMYork96,GTabacchi02,Atlas06,ANiklasson21,JGoff23} From this perspective, flexible-charge equilibration models can be viewed within the framework of conceptual DFT, \cite{Parr89,Geerlings2020} using a constrained, coarse-grained representations of the electron density. Rather than adopting a particular functional form, we will formulate the theory for a broad class of nonlinear energy functionals. This framework includes conventional quadratic charge-equilibration methods, higher-order nonlinear models with atom-centered multipole descriptions, and various machine-learned charge-dependent energy functionals.

The nonlinear charge-dependent energy functional that defines the general flexible-charge model considered in this article, including a net-charge constraint enforced by a Lagrange multiplier $\mu$, is given by
\begin{equation}\label{eq:E_Ansatz}
E({\bf q},\mu) = E_{\rm S}({\bf q}) +\frac{1}{2}{\bf q}^{\rm T}{\bf C}{\bf q} +\mu({\bf 1}^{\rm T}{\bf q}-Q_{\rm tot}).
\end{equation}
In this constrained energy functional (or Lagrangian), ${\bf q}\equiv{q_{ilm}}$ denotes the vector of atom-centered multipoles of the electron-density difference relative to a superposition of isolated atomic electron densities, with atom $i$ located at position ${\bf R}_i$. \cite{ANiklasson21,JGoff23} For example, for each atom $i$, the $l=0$ terms represent the net partial monopole charge (or occupation), while the $l=1$ terms represent the three dipole components. No particular assumption is made about the radial distributions associated with these multipole charge distributions. When referring to the atom-projected net electron density in the flexible-charge model above, we use the terms such as electronic degrees of freedom, electronic multipoles, or more simply charges, interchangeably.

The short-range contribution, $E_{\rm S}({\bf q})\equiv E_{\rm S}({\bf R},{\bf q})$, is an unspecified nonlinear electronic energy functional of the partial multipoles ${\bf q}$ and the atomic positions ${\bf R} \equiv \{{\bf R}_i\}$. The matrix ${\bf C} \equiv {\bf C}({\bf R}) \equiv \{C_{ij}^{(lm,l’m’)}\}$ describes the long-range electrostatic interactions between multipoles on atoms $i$ and $j$. The interaction matrix ${\bf C}$ does not include the onsite ($i = j$) interactions, and no particular radial representation of the multipole densities is assumed. Onsite interactions are thus assumed to be accounted for in $E_{\rm S}({\bf q})$. The vector ${\bf 1}^{\rm T} \equiv [1,1,\ldots,1,0,0,\ldots,0]$ contains unit entries for the monopole ($l = 0$) degrees of freedom and zeros for all higher-order multipoles. The constraint ${\bf 1}^{\rm T}{\bf q} = Q_{\rm tot}$ therefore enforces conservation of the total net partial charge, $Q_{\rm tot}$, with the Lagrange multiplier, $\mu$, playing the role of the chemical potential for the net partial charge or occupation.

The electronic-energy ansatz in Eq.\ (\ref{eq:E_Ansatz}) can be viewed as a coarse-grained form of approximate or conceptual DFT for describing the charge-dependent energy of an atomistic system. Apart from the multipolar representation of the net partial charge density, the framework is deliberately general and is based primarily on a separation of the electronic energy into short-range and long-range contributions or, more generally, into an unspecified, short-range, electronic energy functional and an explicit physics-motivated, long-range interaction term. The unspecified functional $E_{\rm S}({\bf q})$ may, for example, be represented by a conventional quadratic QEq model based on atom-specific electronegativities and chemical hardness parameters, by a higher-order nonlinear multipolar expansion, or by a general machine-learned charge-dependent energy functional represented by a deep neural network. The second, long-range physics-motivated term, which here describes the electrostatic energy, could also include other known charge-dependent interactions, such as dispersion contributions. \cite{Giese2007,Tkatchenko2009,Tkatchenko2012,Ambrosetti2016,Caldeweyher2017,Caldeweyher2019} 
Such interactions, however, are not considered in this article.

The DFT-based ansatz in Eq.\ (\ref{eq:E_Ansatz}) differs from MLIPs in which atomic charges are first predicted from local atomic environments using neural networks or kernel-ridge regression and then used to evaluate additional long-range energy and force contributions. \cite{JBehler11,OUnke19,RZubatyuk21,ZSong2024} It also differs from approaches in which, for example, a second-order global charge-equilibration model determines atomic charges that are subsequently used as additional descriptors for machine-learned energy corrections. \cite{TWKo20,BFuchs2025,DAnstine2025} In contrast, the atomic charges and multipoles in our formulation are determined self-consistently by minimizing the complete electronic energy functional, including both short- and long-range contributions, in direct analogy with the variational principle of Hohenberg–Kohn DFT. \cite{SGoedecker15,Faraji2017,RahmatizadKhajehpasha2022, Staacke2022,JGoff23,Shaidu2024, LiNiklasson25,Vondrak2025,Vondrak2026,RGrove26} However, while most existing variational charge-equilibration models are restricted to quadratic charge dependence, the present framework account also for general nonlinear local electronic energy functionals, including higher-order charge and multipolar contributions.

The interatomic Born-Oppenheimer potential-energy surface, $U_{\rm BO}({\bf R})$, is obtained by evaluating the complete energy functional, $E({\bf q},\mu)$, at its constrained minimum, corresponding to the equilibrated electronic state, where
\begin{equation}
U_{\rm BO}({\bf R}) = V({\bf R}) + \Big\{ E({\bf q},\mu)~ \Big\vert ~\partial E/\partial {\bf q}={\bf 0},~ \partial E/\partial \mu=0 \Big\}.
\end{equation}
Here $V({\bf R})$ is a charge-independent potential-energy term that may include, for example, ion-ion repulsion or charge-independent formulations of van der Waals interaction terms. More general charge-dependent dispersion interactions would require additional long-range, physics-based terms in the electronic energy functional and are not considered in this article.

The equilibrated ground state, $[{\bf q}_0,\mu_0]$, determining the Born-Oppenheimer potential, $U_{\rm BO}({\bf R})=V({\bf R})+E({\bf q}_0,\mu_0)$, is given by the solution of the nonlinear system of equations,
\begin{equation} \label{Eq:NonLinSolve}
\begin{bmatrix}
{\bf C} & {\bf 1}\\
{\bf 1}^{\rm T} & 0
\end{bmatrix}
\begin{bmatrix}
{\bf q}\\
\mu
\end{bmatrix}
=
\begin{bmatrix}
-\nabla_q E_{\rm S}({\bf q})\\
Q_{\rm tot}
\end{bmatrix}.
\end{equation}
This nonlinear system must be solved iteratively to find a self-consistent solution, for example, using Newton’s method. In the special case of conventional second-order QEq models, the system becomes linear. However, even in this case an iterative solver is typically needed, particularly for large periodic systems, where each iteration requires an Ewald summation. 

Once the Born-Oppenheimer potential has been determined, the molecular trajectories of an MD simulation can be generated by integrating Newton’s equations of motion,
\begin{equation}
M_i{\bf \ddot R}_i =
-\frac{\partial U_{\rm BO}({\bf R})}{\partial {\bf R}_i},
\end{equation}
using, for example, the leapfrog velocity Verlet algorithm.

A charge-dependent MLIP that includes long-range charge response can be constructed using the equilibrated Born-Oppenheimer potential, $U_{\rm BO}({\bf R})$, defined above. The charge-independent term, $V({\bf R})$, and the short-range charge-dependent functional, $E_{\rm S}({\bf q})$, can be parameterized using, for example, neural networks or kernel ridge regression trained on high-quality reference data set. \cite{TWKo20,JGoff23,LiNiklasson25} For a general non-linear electronic energy functional this training and regularization can be quite involved and must be performed through the equilibrated Born-Oppenheimer potential. The details of such machine-learning techniques will not be discussed in this article.

The Born-Oppenheimer MD approach described above, including long-range charge-relaxation effects, may appear fairly straightforward, at least in principle. In practice, however, the equilibrated ground-state solution defining $U_{\rm BO}({\bf R})$ must be well converged at every time step. Otherwise, computationally expensive calculations involving nonzero $\sum_k\big(\partial E/\partial {\bf q}_k\big) \big(\partial {\bf q}_k/\partial {\bf R}_i\big)$  terms must be performed.
If these terms are not included, the interatomic forces may not be sufficiently conservative, potentially losing the long-term stability of the dynamics or producing a systematic energy drift that could invalidate the MD simulation. 
Shadow MD has been developed to alleviate these problems, both by reducing the computational cost of the electronic equilibration and by improving the long-term stability of the dynamics.

\section{Shadow Molecular Dynamics for Flexible-Charge Equilibration Models} \label{sec:ShadowMD}

The main computational bottleneck of Born-Oppenheimer MD arises from the iterative solution of the constrained nonlinear optimization problem in Eq.\ (\ref{Eq:NonLinSolve}), which is required at every time step. The central idea of shadow MD is to avoid this repeated nonlinear optimization without significantly compromising the accuracy of the potential-energy surface or losing the conservative nature of the interatomic forces needed to ensure long-term stability.

To achieve this, we construct a nearby \emph{shadow} Born-Oppenheimer potential-energy surface that can be evaluated much more efficiently while closely tracking the corresponding tightly converged reference potential throughout the simulation. Importantly, shadow MD does not approximate the interatomic potential or forces directly. Instead, it introduces an approximate electronic energy functional whose exact equilibration with respect to the charges and multipoles defines the shadow potential. The interatomic forces are then obtained as exact derivatives of this potential and are therefore conservative. Molecular trajectories consequently evolve on a well-defined shadow Born-Oppenheimer potential-energy surface that provides a stable and accurate approximation to the reference Born-Oppenheimer dynamics. This construction is independent of the specific functional form or parameterization of the underlying short-range flexible-charge energy, $E_S({\bf q})$. Whether this energy is quadratic, highly nonlinear, or represented by a neural network, the shadow MD formulation presented below follows from the same variational principles.

\subsection{Shadow Potential}

In shadow MD, we replace the constrained, charge-dependent, energy functional (or Lagrangian), $E({\bf q},\mu)$ in Eq.\ (\ref{eq:E_Ansatz}), by a shadow energy functional, ${\cal E}({\bf q},{\bf n},\mu)$, obtained from a combined linearization and second-order expansion of $E({\bf q},\mu)$ in Eq.\ (\ref{eq:E_Ansatz}) around an approximate ground-state solution ${\bf n}\approx{\bf q}_0$. This shadow energy functional is defined as
\begin{align}
    {\cal E}({\bf q},{\bf n},\mu) =& E_{\rm S}({\bf n}) + \big({\bf q}-{\bf n}\big)^{\rm T} E'_{\rm S}({\bf n}) \nonumber \\
    & + \frac{1}{2} \big({\bf q}- {\bf n}\big)^{\rm T}E''_{\rm S}({\bf n}) \big({\bf q}- {\bf n}\big) \nonumber \\ 
    & \quad + \frac{1}{2} \big(2{\bf q}-{\bf n}\big)^{\rm T}{\bf C} {\bf n} + \mu\big({\bf 1}^{\rm T}{\bf q} - Q_{\rm tot} \big).
\end{align}
Here, $E'_{\rm S}({\bf n}) \equiv \nabla_nE_{\rm S}({\bf n})$ is the multipole gradient vector, while
$E''_{\rm S}({\bf n}) \equiv {\rm diag}\big[{\bf H}_{\rm S}({\bf n})\big]$ is the diagonal part of the short-range (S) Hessian, ${\bf H}_{\rm S}({\bf n})$, with matrix elements,
\begin{align}
\quad \big\{{\bf H}_{\rm S}({\bf n})\big\}_{ij}=\frac{\partial^2E_{\rm S}({\bf q})}{\partial q_i\partial q_j}\Big \vert_{{\bf q} = {\bf n}}.
\end{align}
Thus, the local diagonal curvature of the short-range energy is kept to second order with respect to ${\bf n}$, whereas the remaining interactions are linearized around ${\bf n}$. The second-order term with the diagonal part of the short-range Hessian is needed to provide a well-defined solution for the equilibrated charges or multipoles. Choosing the diagonal part makes the equilibration inexpensive. However, there is no unique form of the shadow energy functional and other expansions are also possible. 

The corresponding ${\bf R}$- and ${\bf n}$-dependent shadow Born-Oppenheimer potential, ${\cal U}_{\rm BO}({\bf R},{\bf n})$, is defined by the exact equilibration of this approximate shadow energy functional, i.e.\
\begin{align}
{\cal U}_{\rm BO}({\bf R},{\bf n})=&
V({\bf R}) + \Big\{
{\cal E}({\bf q},{\bf n},\mu)
~\Big\vert~
\partial{\cal E}/\partial{\bf q}={\bf 0},\nonumber  \\
& ~~~~~~~~~~~~~~~~~~~~~~~~\partial{\cal E}/\partial\mu=0 \Big\},
\end{align}
where $V({\bf R})$ is the charge-independent contribution to the potential energy. For a given ${\bf n}$, the equilibrated shadow charge distribution, ${\bf q}_0[{\bf n}]$, and the corresponding Lagrange multiplier, $\mu_0[{\bf n}]$, are obtained from the quasi-diagonal linear system of equations,
\begin{equation}\label{Eq:q(n)}
\begin{bmatrix}
E''_{\rm S}({\bf n}) & {\bf 1}\\
{\bf 1}^{\rm T} & 0
\end{bmatrix}
\begin{bmatrix}
{\bf q}_0[{\bf n}]\\
\mu_0[{\bf n}]
\end{bmatrix}
=
\begin{bmatrix}
E''_{\rm S}({\bf n}){\bf n}
-E'_{\rm S}({\bf n})
-{\bf C}{\bf n}\\
Q{\rm tot}
\end{bmatrix}.
\end{equation}
The system matrix can be expressed as a rank-2 update of a diagonal matrix and can therefore be inverted analytically in a single step using the Woodbury formula. \cite{LiNiklasson25,RGrove26} Alternatively, the same solution can be obtained using a Schur-complement decomposition. The evaluation of the shadow potential is therefore straightforward. Because the nonlinear terms have been partially linearized, no iterative charge-equilibration procedure is required. The dominant computational cost is a single Coulomb summation in ${\bf C}{\bf n}$, which needs to be performed, for example, with the Ewald summation method for periodic systems.  Problems associated with incomplete iterative convergence are therefore avoided and the shadow electronic state satisfies the stationarity conditions exactly, i.e.\
\begin{equation}
\frac{\partial {\cal E}}{\partial {\bf q}} = {\bf 0},
\qquad
\frac{\partial {\cal E}}{\partial \mu} = 0.
\end{equation}
at ${\bf q}={\bf q}_0[{\bf n}]$ and $\mu=\mu_0[{\bf n}]$.
Although the quasi-diagonal system matrix is local, nonlocal interactions are accounted for through the Coulomb summation, ${\bf C}{\bf n}$.

The shadow Born-Oppenheimer potential can be used to generate the interatomic forces and drive a molecular dynamics. Because $\big(\partial {\cal E}/\partial {\bf q} \big)\big \vert_{{\bf q} = {\bf q}_0[{\bf n}]} = {\bf 0}$, without any convergence errors, the problems with inaccurate non-conservative forces in regular Born-Oppenheimer MD can be avoided. 
The main issue is instead how we deal with the approximate ground-state density ${\bf n}$ as the atoms are moving. We need to keep ${\bf n}$ close to the exact ground state ${\bf q}_0$ and we need to avoid force terms that include $\big(\partial {\cal E}/\partial {\bf n}_k \big)\big( \partial {\bf n}_k/\partial {\bf R}_i\big)$, which in general, would be expensive to calculate. Extended Lagrangian Born-Oppenheimer molecular dynamics can be used to solve these problems. \cite{ANiklasson08,ANiklasson21b}

\subsection{Extended Lagrangian}

The accuracy of the shadow potential, ${\cal U}_{\rm BO}({\bf R},{\bf n})$, depends on how closely ${\bf n}$ approximates the exact, regular Born-Oppenheimer ground state, ${\bf q}_0$. This error can be estimated by the residual $\vert{\bf q}_0[{\bf n}]-{\bf n}\vert$, where ${\bf q}_0[{\bf n}]$ is the ${\bf n}$-dependent ground state of the shadow potential determined by Eq.\ (\ref{Eq:q(n)}).
The residual vanishes at the exact regular ground state density with the self-consistent solution ${\bf n}={\bf q}_0$. To keep ${\bf n}$ close to the ground state throughout an MD simulation, we introduce ${\bf n}(t)$ and ${\bf \dot n}(t)$ as dynamical variables, in addition to the nuclear dynamical variables ${\bf R}(t)$ and ${\bf \dot R}(t)$. This can be achieved using an extended-Lagrangian formalism, \cite{HCAndersen80,MParrinello80,SNose84,RCar85,ANiklasson08,ANiklasson21b} where the extended electronic variables, ${\bf n}(t)$, evolve through a harmonic oscillator centered around the instantaneous ground-state solution. By introducing ${\bf n}(t)$ as independent dynamical variables within an extended-Lagrangian formalism, we also avoid the problem of non-conservative forces. Since ${\bf n}(t)$ and the nuclear coordinates ${\bf R}(t)$ are independent dynamical variables, the interatomic forces are evaluated at fixed ${\bf n}$, and no force terms involving partial derivatives $\partial {\bf n}/\partial {\bf R}_i$ appear, i.e.\ contributions of the form $\sum_k \big(\partial {\cal E}/\partial n_k\big)\big(\partial n_k/\partial {\bf R}_i\big)$ are absent.

Using the notation
\begin{equation}
{\bf c}_0[{\bf x}] \equiv
\begin{bmatrix}
{\bf q}_0[{\bf n}]\\
\mu_0[{\bf n}]
\end{bmatrix},
\qquad
{\bf x} \equiv
\begin{bmatrix}
{\bf n}\\
\nu
\end{bmatrix}
\approx
{\bf c}_0 \equiv
\begin{bmatrix}
{\bf q}_0\\
\mu_0
\end{bmatrix},
\end{equation}
we define the extended Lagrangian as
\begin{align}
{\cal L}({\bf R},{\bf \dot R},{\bf x},{\bf \dot x})
=&
\frac{1}{2}\sum_i M_i\vert{\bf \dot R}_i\vert^2
-{\cal U}_{\rm BO}({\bf R},{\bf n})
+\frac{1}{2}m{\bf \dot x}^{\rm T}{\bf \dot x}
\nonumber\\
&\quad
-\frac{1}{2}m\omega^2
\big({\bf c}_0[{\bf x}]-{\bf x}\big)^{\rm T}
{\bf T}
\big({\bf c}_0[{\bf x}]-{\bf x}\big).
\label{eq:XL}
\end{align}
The first two terms on the right-hand side are the kinetic energy of the nuclei and the ${\bf n}$-dependent shadow Born-Oppenheimer potential. The last two terms describe the extended harmonic oscillator that drives the dynamical electronic degrees of freedom, ${\bf x}(t) \equiv [{\bf n}(t), \nu(t)]^{\rm T}$, using a harmonic well that is centered around the ground-state solution, ${\bf c}_0$.
Here, $m$ is a fictitious mass parameter and $\omega$ is the frequency of the extended harmonic oscillator. As long as the electronic frequency, $\omega$, is high enough compared to the nuclear motion, the electronic degrees of freedom, ${\bf n}(t)$, will stay close to the equilibrated ground state such that the shadow potential remains close to the exact regular Born-Oppenheimer potential.

The matrix ${\bf T}={\bf K}^{\rm T}{\bf K}$ is a symmetric positive-definite metric tensor, where the kernel ${\bf K}={\bf J}^{-1}$ is the inverse of the Jacobian ${\bf J}$ of the residual function, ${\bf f}({\bf x}) = {\bf c}_0[{\bf x}]-{\bf x}$, which can be defined as
\begin{equation}
{\bf J}
\begin{bmatrix}
\delta{\bf n}\\
\delta\nu
\end{bmatrix}
= \delta
\begin{bmatrix}
{\bf q}_0[{\bf n}]-{\bf n}\\
\mu_0[{\bf n}]-\nu
\end{bmatrix}, \label{Eq:J_Def1}
\end{equation}
or, equivalently, as
\begin{equation}
{\bf J}  = \frac{\partial {\bf f}({\bf x})}{\partial {\bf x}} =
\begin{bmatrix}
\displaystyle
\partial{\bf q}_0[{\bf n}]\big/\partial{\bf n}-{\bf I}
& \displaystyle \partial{\bf q}_0[{\bf n}]\big/\partial\nu\\
\displaystyle \partial\mu_0[{\bf n}]\big/\partial{\bf n}
& \displaystyle \partial\mu_0[{\bf n}]\big/\partial\nu-1
\end{bmatrix}.
\label{Eq:J_Def2}
\end{equation}
Notice that in our formulation, ${\bf q}_0[{\bf n}]$ and $\mu_0[{\bf n}]$ depend only on ${\bf n}$ and not on $\nu$, thus
\begin{equation}
\frac{\partial{\bf q}_0[{\bf n}]}{\partial\nu}={\bf 0},
\qquad
\frac{\partial\mu_0[{\bf n}]}{\partial\nu}=0.
\end{equation}

To determine the shadow MD generated by the extended Lagrangian above, we first need to derive the corresponding equations of motion.

\subsection{Equations of Motion}

The equations of motion for shadow MD follow from the extended Lagrangian in Eq.\ (\ref{eq:XL}) through the Euler-Lagrange equations,
\begin{align}
\frac{d}{dt}\left(\frac{\partial{\cal L}}{\partial{\bf \dot R}_i}\right)
&= \frac{\partial{\cal L}}{\partial{\bf R}_i}, \qquad
\frac{d}{dt}\left(\frac{\partial{\cal L}}{\partial{\bf \dot x}}\right)= \frac{\partial{\cal L}}{\partial{\bf x}}.
\end{align}
The equations of motion are derived in an adiabatic limit. In analogy with the Born-Oppenheimer approximation, \cite{marx_hutter_2009} we assume that the extended dynamical electronic degrees of freedom, ${\bf n}(t)$ and ${\bf \dot n}(t)$, evolve on a faster time scale than the heavier and slower nuclear degrees of freedom. \cite{ANiklasson21b}
 The separation appears when the auxiliary electronic degrees of freedom have a small mass and move rapidly in the extended harmonic oscillator. This adiabatic limit can be obtained by letting $m\rightarrow 0$ and $\omega\rightarrow\infty$, while keeping $m\omega$ constant. We further assume that
$|{\bf c}_0[{\bf x}]-{\bf x}|\propto\omega^{-2}$ and that $|\partial {\cal U}_{\rm BO}\big / \partial {\bf n} | \propto \omega^{-2}$. \cite{ANiklasson21b}
In this limit, the equations of motion become
\begin{align}
M_i{\bf \ddot R}_i
&= -\left. \frac{\partial{\cal U}_{\rm BO}({\bf R},{\bf n})}
{\partial{\bf R}_i} \right|_{\bf n},\\
{\bf J}{\bf \ddot x}
&= -\omega^2\big({\bf c}_0[{\bf x}]-{\bf x}\big).
\label{Eq:AdiabaticEOM}
\end{align}
The interatomic forces are calculated from the partial derivatives with respect to the nuclear coordinates under constant ${\bf n}$, because ${\bf n}(t)$ are dynamical variables. 
In the adiabatic limit we ignore all terms $\propto |{\bf q}_0[{\bf n}]-{\bf n}|^2$, which therefore also can be avoided in the force evaluations.

In the adiabatic limit the Jacobian, ${\bf J}$, is given (see Appendix) by
\begin{equation}
{\bf J}=
- \begin{bmatrix}
E''_{\rm S}({\bf n}) & {\bf 1}\\
{\bf 1}^{\rm T} & 0
\end{bmatrix}^{-1}
\begin{bmatrix}
{\bf C}+{\bf H}_{\rm S}({\bf n}) & {\bf 1}\\
{\bf 1}^{\rm T} & 0
\end{bmatrix},
\label{Eq:Adiabatic_J}
\end{equation}
which can also be viewed as a fixed-point approximation to the Jacobian at the exact equilibrated state, i.e.\ evaluated under the condition that ${\bf c}_0[{\bf x}]={\bf x}$.

Combining this adiabatic Jacobian with the equations for the equilibrated ${\bf n}$-dependent electronic variables, ${\bf c}_0[{\bf x}]$, in Eq.\ (\ref{Eq:q(n)}), gives the final equations of motion defining our shadow MD for general nonlinear flexible-charge models,
\begin{align}
M_i {\bf \ddot R}_i
&= -\left. \frac{\partial{\cal U}_{\rm BO}({\bf R},{\bf n})} {\partial{\bf R}_i} \right|_{\bf n}, \label{Eq:d2Rdt2} \\
\begin{bmatrix}
{\bf C}+ {\bf H}_{\rm S}({\bf n}) & {\bf 1}\\
{\bf 1}^{\rm T} & 0
\end{bmatrix}
\begin{bmatrix}
{\bf \ddot n}\\
\ddot\nu
\end{bmatrix}
&= -\omega^2
\begin{bmatrix}
{\bf C}{\bf n}+\nabla_nE_{\rm S}({\bf n})+\nu{\bf 1}\\
{\bf 1}^{\rm T}{\bf n}-Q_{\rm tot}
\end{bmatrix}.
\label{Eq:ChargeAcc}
\end{align}
These equations together with their equivalent but more compact form in Eqs.\ (\ref{Eq:d2Rdt2_b}) and (\ref{Eq:d2ndt2_b}) is the central result of this paper. Details of their derivation are given in the Appendix and their derivation from an alternative extended Lagrangian together with a stability analysis are given in Section\ \ref{Sec:AltAnalysis}. These equations govern the shadow dynamics and are exact in continuous time as the integration time step $\delta t \rightarrow 0$ while $\delta t \times \omega = {\rm constant}$. 

Of particular importance is that the atomic forces in Eq.\ (\ref{Eq:d2Rdt2}) are evaluated as partial derivatives of the shadow potential with respect to the atomic coordinates at fixed ${\bf n}$. Because ${\bf n}(t)$ is an independent dynamical variable of the extended system, no derivatives of ${\bf n}$ with respect to the nuclear coordinates are required. This provides a significant simplification in the evaluation of the interatomic forces.

The equations of motion are quite general and can be applied to conventional second-order charge-equilibration models, higher-order nonlinear flexible-charge models with multipole interactions, and charge-dependent energy functionals represented by machine-learned models such as deep neural networks. 

The first equation is Newton’s equation of motion for the nuclei and can be integrated, for example, using the leapfrog velocity-Verlet scheme, as described below. The second equation describes the dynamics of the extended electronic degrees of freedom driven by a harmonic oscillator. The equation determines the charge or multipole acceleration, ${\bf \ddot n}(t)$, implicitly through a linear system of equations and captures the long-range response dynamically.  This system can be solved, for example, using the generalized minimal residual (GMRES) method, \cite{saad1986gmres,golub2013matrix,saad2003iterative} replacing the low-rank kernel expansions used in earlier formulations. \cite{ANiklasson21,ANiklasson21b,JGoff23,RStanton25,Li2025,RGrove26}.

Because ${\bf q}_0[{\bf n}]$ and $\mu_0[{\bf n}]$ are determined by ${\bf n}(t)$ and are independent of $\nu$ in Eq.\ (\ref{Eq:q(n)}), only ${\bf n}(t)$ needs to be integrated in the shadow MD. Moreover, in the harmonic equation of motion, Eq.\ (\ref{Eq:ChargeAcc}), the term $\nu{\bf 1}$ represents only a uniform shift of the electrostatic potential and does not affect the charge acceleration ${\bf \ddot n}$. We may therefore set $\nu=0$ without loss of generality.
If we further assume the charge-conservation condition
\begin{equation}
{\bf 1}^{\rm T}{\bf n}=Q_{\rm tot}, \label{eq:NetQ}
\end{equation}
is fulfilled, then ${\bf 1}^{\rm T}{\bf \ddot n}=0$, and the equation for ${\bf \ddot n}(t)$ can be reduced to
\begin{equation}\label{eq:PCG}
\left[ {\bf C}+{\bf H}_{\rm S}({\bf n})\right] {\bf \ddot n}=
-\omega^2 \left[ {\bf C}{\bf n}+\nabla_nE_{\rm S}({\bf n}) \right],
\end{equation}
subject to the constraint
\begin{equation}
{\bf 1}^{\rm T}{\bf \ddot n}=0. \label{eq:Q_constr}
\end{equation}
If the initial charges, ${\bf n}(t_0)$, satisfy the net charge constraint in Eq.\ (\ref{eq:NetQ}) and the calculated accelerations remain charge neutral, as enforced by Eq.\ (\ref{eq:Q_constr}), the net charge constraint is preserved throughout the dynamics when using a charge-conserving integration scheme. The constraint in Eq.\ (\ref{eq:Q_constr}) can be enforced using a projected preconditioned conjugate-gradient (CG) method. \cite{MRHestenes52,Nocedal2006} A suitable projected preconditioner is
\begin{equation}\label{eq:G}
{\bf M}= {\bf G} -
\frac{{\bf G}{\bf 1}{\bf 1}^{\rm T}{\bf G}}{{\bf 1}^{\rm T}{\bf G}{\bf 1}},
\qquad
{\bf G} \approx \left( {\bf C}+ {\bf H}_{\rm S}({\bf n})\right)^{-1}.
\end{equation}
The corresponding projected preconditioned CG algorithm for solving Eq.\ (\ref{eq:PCG}) is described in Alg.\ \ref{Alg:PCG} below.

Notice, that the CG method requires that the system matrix, ${\bf C}+{\bf H}_{\rm S}({\bf n})$, remains positive definite. For general non-linear functionals ${\bf C}+{\bf H}_{\rm S}({\bf n})$ could, at least in principle, become indefinite, which would require a preconditioned GMRES solver. However, such examples are beyond the scope of this article.

The shadow potential, ${\cal U}_{\rm BO}({\bf R},{\bf n})$, remains accurate over a relatively broad region around the partial-linearization point defined by ${\bf n}(t)$ \cite{RStanton25,RGrove26}. The linear solution for ${\bf \ddot n}(t)$ therefore does not need to be tightly converged, provided that the total charge is conserved. As we will demonstrate in the examples below, this property can substantially reduce the computational cost without significantly compromising the long-term stability of the dynamics or the accuracy of the shadow potential-energy surface.
In fact, the iterative solution of ${\bf \ddot n}$ can often be avoided altogether by applying the projected preconditioner, ${\bf M}$, directly as a fixed approximate inverse, with
\begin{equation}\label{eq:FixedPrecond}
{\bf \ddot n} \approx 
-\omega^2{\bf M}
\big[ {\bf C}{\bf n}+\nabla_n E_{\rm S}({\bf n}) \big].
\end{equation}
If necessary, the preconditioner, ${\bf M}$, can be updated at relatively little additional cost at suitable intervals, for example every 100 or 1000 MD time steps. Such updates may be required to maintain stability and accuracy as the atomic configuration evolves significantly over time. Averaged over a longer trajectory, however, the computational cost of updating the preconditioner is negligible. Alternatively, the fixed preconditioning approximation in Eq.\ (\ref{eq:FixedPrecond}) can simply be used as the initial step of a preconditioned CG algorithm to solve Eq.\ (\ref{eq:PCG}).

The three forms of the charge-dependent equations of motion above, Eqs.\ (\ref{Eq:ChargeAcc}), (\ref{eq:PCG}), and (\ref{eq:FixedPrecond}), provide the basis for three different strategies for determining the charge or multipole acceleration, ${\bf \ddot n}(t)$. Equation (\ref{Eq:ChargeAcc}) can be solved using preconditioned GMRES; \cite{saad1986gmres} Eq.\ (\ref{eq:PCG}) can be solved using projected preconditioned CG; \cite{MRHestenes52,Nocedal2006} and Eq.\ (\ref{eq:FixedPrecond}) provides an explicit approximation based on a fixed preconditioner. The last approach is particularly attractive because its computational cost can approach that of an electrostatic model with predefined fixed charges or multipoles. If the fixed-preconditioner approximation provides sufficient accuracy and stability, the additional cost of treating self-consistent flexible charges or multipoles can therefore become quite small. In previous formulations of shadow MD, this limit has been difficult to achieve in practice, and a small number of low-rank kernel or Krylov-subspace corrections have typically been required. \cite{JGoff23,LiNiklasson25,RStanton25,Kaymak2025GPU,RGrove26}

\subsection{Integrating the Equations of Motion}

Once the atomic and charge or multipole accelerations, ${\bf \ddot R}(t)$ and ${\bf \ddot n}(t)$, have been determined, we can integrate the equations of motion. We can do this using a modified leapfrog velocity-Verlet scheme,
\begin{align}
&{\bf \dot R}(t + \delta t/2) = {\bf \dot R}(t)  + \frac{1}{2}\delta t {\bf \ddot R}(t)\\
& {\bf R}(t+ \delta t) = {\bf R}(t) + \delta t {\bf \dot R}(t + \delta t/2)\\
&{\bf n}(t+\delta t) = 2{\bf n}(t) - {\bf n}(t-\delta t) + \delta t^2 {\bf \ddot n}(t) + {\rm Diss}({\bf n},\delta t^{2k+1})\label{Eq:Verlet_n}\\
&{\bf \dot R}(t + \delta t) = {\bf \dot R}(t + \delta t/2)+ \frac{1}{2}\delta t {\bf \ddot R}(t + \delta t).
\end{align}
In the initial time step we use a regular Born-Oppenheimer solution, such that ${\bf x}(t_0) \equiv [{\bf n}(t_0),\nu(t_0)]^{\rm T}= {\bf c}_0(t_0) = [{\bf q}_0(t_0),\mu_0(t_0)]^{\rm T}$. The term ${\rm Diss}({\bf n},\delta t^{2k+1})$ is a weak dissipative force term that breaks time reversibility in some higher odd order of the integration time step, $\delta t$, which includes between 5 to 7 charge or multipole-vectors from previous time steps. \cite{ANiklasson09, PSteneteg10, GZheng11} This dissipation or friction term is included to avoid an accumulation of numerical noise and to keep the evolution of ${\bf n}(t)$ synchronized with ${\bf R}(t)$. \cite{ANiklasson21b} The dissipation can be seen as part of a Langevin-like dynamics where a dissipative force balances the intrinsic numerical noise apparent in the simulation and improves the adiabatic separation between the nuclear and electronic degrees of freedom. Alternative approaches for the integration are also possible. \cite{Albaugh15,VVitale17,ILeven19}

Integrating the equations of motion for the extended electronic degrees of freedom in Eq.\ (\ref{Eq:Verlet_n}) requires the evaluation of the charge or multipole accelerations, ${\bf \ddot n}(t)$, in Eq.\ (\ref{Eq:ChargeAcc}). To determine the charge acceleration we therefore need to solve for ${\bf \ddot n}(t)$ as given in Eq.\ (\ref{Eq:ChargeAcc}), (\ref{eq:PCG}) or (\ref{eq:FixedPrecond}). In general, this can be achieved at low cost, because high accuracy is not required as long as stability is maintained, and the preconditioner needs to be updated only infrequently.

\begin{figure}[tb]
\centering
\refstepcounter{myalgorithm}
\label{Alg:PCG}
\begin{minipage}{0.98\columnwidth}
\smallskip
\noindent\textbf{Algorithm \themyalgorithm.}
Projected preconditioned conjugate-gradient solver. \cite{Nocedal2006} Solves $\mathbf A\mathbf x=\mathbf b$ subject to the constraint $\mathbf 1^{\rm T}\mathbf x=0$. The convergence criterion can alternatively be based on the root-mean-square preconditioned residual or its relative reduction.
\medskip
\begin{algorithmic}
\Require Symmetric matrix $\mathbf A$,
right-hand side $\mathbf b$,
initial guess $\mathbf x_0$ satisfying
$\mathbf 1^{\rm T}\mathbf x_0=0$,
preconditioner $\mathbf M$ as in Eq.~(\ref{eq:G}),
and convergence tolerance $\epsilon$.
\State $\mathbf x=\mathbf x_0$,
       $\mathbf r=\mathbf b-\mathbf A\mathbf x$
\State $\mathbf z=\mathbf M\mathbf r$,
       $\mathbf p=\mathbf z$,
       $\sigma=\mathbf r^{\rm T}\mathbf z$
\While{not converged}
    \State $\mathbf q=\mathbf A\mathbf p$
    \State $\alpha=\sigma/(\mathbf p^{\rm T}\mathbf q)$
    \State $\mathbf x=\mathbf x+\alpha\mathbf p$
    \State $\mathbf r=\mathbf r-\alpha\mathbf q$
    \State $\mathbf z=\mathbf M\mathbf r$
    \If{$\|\mathbf z\|<\epsilon$}
        \State \textbf{break}
    \EndIf
    \State $\sigma_{\rm new}=\mathbf r^{\rm T}\mathbf z$
    \State $\beta=\sigma_{\rm new}/\sigma$
    \State $\mathbf p=\mathbf z+\beta\mathbf p$
    \State $\sigma=\sigma_{\rm new}$
\EndWhile
\State \textbf{return} $\mathbf x$
\end{algorithmic}
\end{minipage}
\end{figure}

 If the projected preconditioned CG method in Alg.\ \ref{Alg:PCG} is used to solve for ${\bf \ddot n}(t)$, we always set the initial guess, ${\bf x}_0 = 0$. The Coulomb potential term ${\bf C}{\bf n}$ required in the initial residual ${\bf r}$ (in the ${\bf A}{\bf x}$ operation) has already been calculated in the solution of the ${\bf q}_0[{\bf n}]$ in Eq.\ (\ref{Eq:q(n)}) and can therefore be reused without extra cost. In each subsequent conjugate gradient iteration an ${\bf A}{\bf p}$ operation is then required in Alg.\ \ref{Alg:PCG}. In our case this includes a Coulomb summation ${\bf C}{\bf p}$, which for periodic boundary conditions can be performed with ${\cal O}(N \log N)$ complexity using the particle-mesh Ewald summation method. \cite{darden1993_pme,Kaymak2025GPU} The  ${\bf A}{\bf p}$ operation also involves an evaluation of ${\bf H}_{\rm S}({\bf n}){\bf p}$. However, we never need to calculate the full Hessian, ${\bf H}_{\rm S}({\bf n})$, which could be expensive. We only need to calculate the Hessian acting on a vector ${\bf p}$, where
\begin{equation}
    {\bf H}_{\rm S}({\bf n}){\bf p}
    = \frac{d}{d\lambda} \nabla_nE_{\rm S}({\bf n} + \lambda {\bf p})\Big\vert_{\lambda = 0}.
\end{equation}
These directional derivatives can be evaluated, for example, with automatic differentiation, at much lower cost. The cost will depend, for example, on details of the parameterization of $E_{\rm S}({\bf n})$ and the non-local range in its charge dependency.

\subsection{Alternative Extended Lagrangian Formulation and Stability Analysis}\label{Sec:AltAnalysis}

Instead of defining the extended Lagrangian as in Eq.\ (\ref{eq:XL}) we can use an alternative formulation leading to the same shadow MD equations of motion as in Eqs.\ (\ref{Eq:d2Rdt2}) and (\ref{Eq:ChargeAcc}) in the adiabatic limit that assumes the separation of timescales between the charge and nuclear degrees of freedom.  In this alternative formulation we define the extended Lagrangian as
\begin{align}
{\cal L}({\bf R},{\bf \dot R},{\bf x},{\bf \dot x})
=&
\frac{1}{2}\sum_i M_i\vert{\bf \dot R}_i\vert^2
-{\cal U}_{\rm BO}({\bf R},{\bf n}) 
 +\frac{1}{2}m\|{\bf \dot x}\|^2\nonumber\\
&~~~-\frac{1}{2}m\omega^2
\|\big({\bf H}({\bf x})\big)^{-1}\nabla_x {E}({\bf x})\|^2,
\label{eq:XL_alt}
\end{align}
where 
\begin{equation}
E({\bf x}) = E_S({\bf n})  + \frac{1}{2} {\bf n}^{\rm T}{\bf C} {\bf n} + \nu({\bf 1}^{\rm T}{\bf n} - Q_{\rm tot} ),
\end{equation}
and
\begin{equation}
    \big \{ {\bf H}({\bf x})\big\}_{ij} = \frac{\partial^2 E({\bf x})}{\partial x_i \partial x_j}
\end{equation}
with
\begin{equation}
     {\bf x} \equiv 
     \begin{bmatrix}
         {\bf n}\\
        \nu
     \end{bmatrix},  \quad \|{\bf a}\|^2 = {\bf a}^{\rm T}{\bf a}.
\end{equation}
Here $\nabla_x E({\bf x})$ is the gradient of the constrained energy expression (or Lagrangian), $E({\bf q},\mu)$ in Eq.\ (\ref{eq:E_Ansatz}), but with ${\bf q}$ and $\mu$ replaced by the dynamical variables ${\bf n}(t)$ and $\nu(t)$ collected in ${\bf x}(t)$.
The inverse-Hessian-scaled square of the gradient of the constrained energy, $\|\big({\bf H}({\bf x})\big)^{-1}\nabla{E}({\bf x})\|^2$, forms the harmonic potential in the extended Lagrangian, which has a minimum for the exact equilibrated charge solution. The corresponding metric tensor in Eq.\ (\ref{eq:XL}) is here given by ${\bf T} = ({\bf H}^{{\rm T}}{\bf H})^{-1}$. Thus ${\bf x}(t)$ in the alternative extended Lagrangian above in Eq.\ (\ref{eq:XL_alt}) will also oscillate around the exact equilibrated ground state just like in Eq.\ (\ref{eq:XL}). The two Lagrangians are different, but in the adiabatic limit, provided ${\bf x}(t)$ remains close to the instantaneous ground state solution, they generate the same equations of motion as in Eqs.\ (\ref{Eq:d2Rdt2}) and (\ref{Eq:ChargeAcc}), which we also can express as
\begin{align}
& M_i {\bf \ddot R}_i
= -\left. \frac{\partial{\cal U}_{\rm BO}({\bf R},{\bf n})} {\partial{\bf R}_i} \right|_{\bf n}, \label{Eq:d2Rdt2_b} \\
&  \big({\bf H}({\bf x})\big) {\bf \ddot x} = - \omega^2 \nabla_x{E}({\bf x}).\label{Eq:d2ndt2_b}
\end{align}
The advantage of this alternative but adiabatically equivalent formulation of the central result of this paper is that it highlights the underlying structure and stability of the extended electronic dynamics. It also facilitates generalizations to alternative representations and more general constrained energy expressions for the shadow MD.

It is easy to demonstrate the harmonic-oscillator equations of motion for the extended electronic degrees of freedom, ${\bf x}(t)$, in the vicinity of the exact equilibrated ground state,
${\bf c}_0 \equiv {\bf c}_0({\bf R})\equiv[{\bf q}_0,\mu_0]^{\rm T}$. Using the local expansion
\begin{equation}
\nabla_x E({\bf x})
\approx
{\bf H}({\bf c}_0) 
\big({\bf x}-{\bf c}_0\big),
\end{equation}
we get
\begin{equation}
{\bf \ddot x}
\approx
-\omega^2\big({\bf x}-{\bf c}_0({\bf R})\big),
\end{equation}
where the explicit ${\bf R}(t)$-dependency on the nuclear coordinates in ${\bf c}_0({\bf R})$ has been included for clarity.
Thus, in the vicinity of the equilibrated solution, where ${\bf x}(t) \approx {\bf c}_0({\bf R})$, the extended electronic dynamics is governed by a stable harmonic oscillator with a constant curvature, $\omega^2$, which is independent of the system-specific charge-response Hessian, ${\bf H}({\bf x})$. All components oscillate with the same frequency, $\omega$, around the exact equilibrated ground-state solution, ${\bf c}_0({\bf R})$, evolving along the propagated molecular trajectory. This  facilitates the control of the adiabatic separation between the charge and nuclear degrees of freedom, for which the highest relevant nuclear frequency, $\Omega_R^{\rm max}$, should remain well below $\omega$.  In shadow MD simulations, integration time steps of the same size as those typically used in regular Born-Oppenheimer MD provide a suitable frequency window for maintaining this adiabatic separation. \cite{ANiklasson21b} However, sudden increases in the nuclear vibrational frequencies, for example, due to exothermic chemical reactions, may lead to an overlap with the frequencies of the extended electronic motion. In this case the adiabatic separation is lost, which could result in energy transfer from the nuclear to the electronic degrees of freedom, potentially causing resonances that would destabilize the shadow MD simulations. 

Apart from alternative extended-Lagrangian formulations, shadow MD can also be constructed using different forms of the shadow potential. \cite{ANiklasson21b} Here the on-site charge-dependent energy diagonal contribution is expanded to second order, while the off-diagonal long-range electrostatic term is expanded to first order. However, other splittings can be used, for example, into block-diagonal and off-block-diagonal contributions. We may also construct higher-level shadow potentials, although at a slightly higher computational cost. For example, the first-level shadow potential requires one additional Coulomb potential summation. \cite{Niklasson2023} Because of these flexibilities, shadow MD is best viewed as a general framework covering a broad range of formulations rather than as a single, uniquely defined method. The same flexibility applies to the different flexible charge-equilibration models that can be represented by the energy functional $E({\bf q},\mu)$ in Eq.\ (\ref{eq:E_Ansatz}).

\section{Model Examples, Advantages and Limitations}
\label{sec:DifferentModels}

Our flexible charge-equilibration Ansatz, based on the energy functional $E({\bf q},\mu)$ in Eq.\ (\ref{eq:E_Ansatz}), represents a fairly general class of nonlinear energy functionals. The short-range energy functional, $E_S({\bf q})$, is conceptual and does not prescribe a specific functional form or the terms to be included. This Ansatz therefore incorporates a number of different flexible-charge models, such as frequently used quadratic charge-equilibration methods, higher-order nonlinear models, atom-centered multipole descriptions, and charge-dependent neural-network functionals. It may also include arbitrary sets of unspecified latent variables, such as potential-fluctuation variables \cite{TVerstraelen13,RStanton25} or even local spin-degrees of freedom apart from the atomic charges or multipoles. 

For MLIPs, it is important that $E_{\rm S}({\bf R},{\bf q})$ is short ranged. Only then can its dimensionality be limited sufficiently to enable efficient training and regularization using precomputed first-principles data. The short-range form also reduces the computational cost of evaluating Hessian-vector products in the iterative Krylov-subspace solvers used to calculate ${\bf \ddot n}(t)$.


To demonstrate our shadow MD framework, we will consider two particular examples: a conventional second-order charge-equilibration model; \cite{Kaymak2025GPU} and a higher-order nonlinear charge-equilibration model. In both cases, we include only the monopole terms of the partial-charge or net occupation. 

Our general formulation of the equations of motion for the extended charge degrees of freedom differs from previous approaches. In particular, it enables the use of off-the-shelf generalized minimal residual (GMRES) \cite{saad1986gmres,golub2013matrix} or preconditioned conjugate-gradient (CG) solvers. Previous formulations relied on a particular low-rank kernel approximation, ${\cal K} \approx {\cal J}^{-1}$, in Eqs.\ (\ref{Eq:AdiabaticEOM}) and (\ref{Eq:Adiabatic_J}), which can be more difficult to implement.

While the new formulation is mathematically the same in the exact integration of the equations of motion, they are different under approximate integration. In particular, the integration of the extended charge degrees of freedom is more robust, allowing computationally efficient approximations. We observe this when we compare the performance with our previous formulation using the same underlying second-order charge-equilibration model in the examples below. \cite{Kaymak2025GPU} In particular, we find that the approximate fixed projected-preconditioning formulation in Eq.\ (\ref{eq:FixedPrecond}) is more stable than the corresponding fixed-kernel approximation, ${\cal K} \approx {\cal J}^{-1}$, in Eq.\ (\ref{Eq:AdiabaticEOM}).

\subsection{Second-Order Flexible-Charge Equilibration Model}\label{QEq_model}

If we approximate $E_{\rm S}({\bf q})$ as a sum of $N$ atom-centered energies, each represented by a second-order expansion in the net partial monopole charge, i.e.,
\begin{equation}
E_{\rm S}({\bf q}) = \sum_{i=1}^N E^0_i + \sum_{i=1}^N \chi_i q_i + \frac{1}{2} \sum_{i = 1}^N u_i q_i^2, \label{Eq:2nd_Order_QEQ}
\end{equation}
we recover a conventional second-order flexible-charge equilibration (QEq) model. \cite{FJVesely77,Gasteiger80,WJMortier86,MSprik88,MSprik90,AKRappe91} This quadratic formulation and its numerous extensions and variations represent the most widely used class of flexible-charge models \cite{DVanBelle92,WSRick94,TAHalgren01,GLamoureaux03,GAKaminsky04,PEMLopes09,PCieplak09,SNaserifar17,JZhifeng19,YLi19,Jensen2023} and will be tested and evaluated in greater detail in the examples below.

Here, the linear coefficients, ${\chi_i}$, correspond to the atomic electronegativities, and ${u_i}$ are the chemical hardness parameters. Note that if ${\bf q}$ denotes the net partial occupation rather than the charge, the electronegativities used here have the opposite sign compared with the conventional definition. Typically, these parameters are specified for each atomic type. However, the model can be extended to include environment-dependent electronegativities, $\chi_i \equiv \chi_i({\bf R})$, and chemical hardness parameters, $u_i \equiv u_i({\bf R})$, which can be trained on large data sets using modern machine-learning methods. \cite{Faraji2017,TWKo20,JGoff23,Shaidu2024,LiNiklasson25}

For the regular QEq model above in Eq.\ (\ref{Eq:2nd_Order_QEQ}), the harmonic-oscillator equations of motion for the extended electronic degrees of freedom in shadow MD are given by
\begin{align}
\begin{bmatrix}
{\bf C}+{\bf U} & {\bf 1}\\
{\bf 1}^{\rm T} & 0
\end{bmatrix}
\begin{bmatrix}
{\bf \ddot n}\\
{\ddot \nu}
\end{bmatrix}
&= -\omega^2
\begin{bmatrix}
{\bf C}{\bf n}+{\boldsymbol \chi} + \nu {\bf 1}\\
{\bf 1}^{\rm T}{\bf n} - Q_{\rm tot}
\end{bmatrix},
\label{Eq:2nd_Order_QEQ_EOMx}
\end{align}
or in its simplified form,
\begin{align}
\begin{bmatrix}
{\bf C}+{\bf U} & {\bf 1}\\
{\bf 1}^{\rm T} & 0
\end{bmatrix}
\begin{bmatrix}
{\bf \ddot n}\\
\lambda
\end{bmatrix}
&= -\omega^2
\begin{bmatrix}
{\bf C}{\bf n}+{\boldsymbol \chi} \\
0
\end{bmatrix}.
\label{Eq:2nd_Order_QEQ_EOM}
\end{align}
Here, ${\bf U}={\rm diag}(\{u_i\}_{i=1}^N)$ is the diagonal matrix of chemical hardness parameters, and ${\boldsymbol \chi}=\{\chi_i\}$ is the vector of atomic electronegativities. The variable $\lambda$ is just a Langrange parameter used to enforce charge neutrality in ${\bf \ddot n}$. 

\subsection{Higher-Order Nonlinear Charge-Equilibration Models}\label{Mth_Order_QEq_Model}

We can generalize the second-order flexible-charge expansion of the atomic energies in $E_{\rm S}({\bf q})$ by including $M$ higher-order terms in the monopole charges, i.e.,
\begin{equation}
E_{\rm S}({\bf q}) = \sum_{i=1}^N E^0_i + \sum_{i=1}^N \chi_i q_i
+ \sum_{i=1}^N \sum_{k=2}^M \frac{1}{k!} u_i^{(k)}q_i^k. \label{Eq:4th_order_QEQ}
\end{equation}
In this case, the harmonic-oscillator equations of motion for the extended electronic degrees of freedom in shadow MD are given by
\begin{align}
\begin{bmatrix}
{\bf C}+E''_{\rm S}({\bf n}) & {\bf 1}\\
{\bf 1}^{\rm T} & 0
\end{bmatrix}
\begin{bmatrix}
{\bf \ddot n}\\
{\ddot \nu}
\end{bmatrix}
&=
-\omega^2
\begin{bmatrix}
{\bf C}{\bf n}+E'_{\rm S}({\bf n}) + \nu{\bf 1}\\
{\bf 1}^{\rm T}{\bf n} - Q_{\rm tot}
\end{bmatrix},
\label{Eq:Mth_Order_QEQx}
\end{align}
or in its simplified form,
\begin{align}
\begin{bmatrix}
{\bf C}+E''_{\rm S}({\bf n}) & {\bf 1}\\
{\bf 1}^{\rm T} & 0
\end{bmatrix}
\begin{bmatrix}
{\bf \ddot n}\\
\lambda
\end{bmatrix}
&=
-\omega^2
\begin{bmatrix}
{\bf C}{\bf n}+E'_{\rm S}({\bf n}) \\
0
\end{bmatrix}.
\label{Eq:Mth_Order_QEQ}
\end{align}
Here,
\begin{equation}
E''_{\rm S}({\bf n})
=
{\rm diag}\left(
\left\{
\sum_{k=2}^M \frac{k(k-1)}{k!}u_i^{(k)}n_i^{k-2}
\right\}_{i=1}^N
\right)
\end{equation}
is the diagonal Hessian matrix of $E_S({\bf q})$ evaluated at ${\bf q}={\bf n}$, and
\begin{equation}
E'_{\rm S}({\bf n})
=
\left[
\frac{\partial E_{\rm S}({\bf n})}{\partial n_1},
\frac{\partial E_{\rm S}({\bf n})}{\partial n_2},
\ldots,
\frac{\partial E_{\rm S}({\bf n})}{\partial n_N}
\right]^{\rm T}
\end{equation}
is the corresponding gradient vector.


In conventional Born-Oppenheimer MD, these higher-order nonlinear charge-equilibration models would require a self-consistent iterative solution at each time step, \cite{MZhang2009,OdaTakahashi2010} making them less attractive for molecular dynamics simulations. The shadow MD formulation avoids this iterative nonlinear optimization by replacing it with the solution of the linear system in Eq.\ (\ref{Eq:Mth_Order_QEQ}).

\subsection{Other Models and Limitations}

The energy functional (or Lagrangian), $E({\bf R},\mu)$ in Eq.\ (\ref{eq:E_Ansatz}), may also accommodate additional hidden or latent variables within the short-range electronic energy functional, $E_{\rm S}({\bf q})$. Such implicit variables could account for intermediate electronic degrees of freedom or additional constraints associated with charge partitioning or fragmentation, as included, for example, in split-charge, capacitor, ACKS2, and related charge-transfer models. \cite{RNistor06,JChen08,TVerstraelen13, Froitzheim2025} 

In a DFT-based formulation, a sufficiently flexible form of $E({\bf q},\mu)$ could, at least in principle, also account for the nonlinear and environment-dependent energy dependence on fractional charges, as considered in ensemble and constrained-search DFT descriptions of interacting subsystems. \cite{SValone2004,Atlas06, Atlas_2021} In the limit where such a subsystem is separated from its environment, the functional should recover the piecewise-linear dependence of the energy as a function of the fractional electron number for the subsystem, together with the derivative discontinuities at integer electron numbers, which is required for an isolated system in exact DFT. \cite{JPerdew1982,WYang2000,Sanches2008} Whether this limiting behavior can be represented by a purely local, smooth electronic energy functional, $E_{\rm S}({\bf R},{\bf q})$, together with a conventional nonlocal multipole interaction alone is less clear and may ultimately limit the ability of the present model to reproduce this exact DFT behavior.

\section{Examples}
\label{sec:Examples}

To evaluate and demonstrate the proposed shadow MD framework for flexible-charge equilibration models, we first consider two representative examples: (A) a fourth-order flexible-charge equilibration model as presented in Sec.\ \ref{Mth_Order_QEq_Model} above \cite{MZhang2009,OdaTakahashi2010} and (B) the more commonly used second-order flexible-charge equilibration model as discussed in Sec.\ \ref{QEq_model} above. In both cases, periodic boxes of liquid water are used as test-bed systems. 

Our first examples are based on microcanonical (NVE) shadow MD simulations. We thereafter examine the behavior of the shadow MD framework in canonical (NVT) simulations using Langevin dynamics. From these Langevin shadow MD simulations we analyze the thermal fluctuations and infrared (IR) spectra. In Langevin dynamics, the additional noise introduced from the thermostat could affect the stability and accuracy of the shadow dynamics, especially in the case of charge-dependent properties such as the IR spectra. 

The objective of these examples is to demonstrate the accuracy, robustness, and computational efficiency of the proposed shadow MD framework. By accuracy, we mean the agreement of the shadow potential-energy surface and the resulting dynamics with those obtained from conventional, tightly converged Born-Oppenheimer MD. Our goal is therefore not to develop or optimize MLIPs for agreement with high-level electronic-structure theory or experimental properties of liquid water. Instead, the water models serve as representative, yet realistic, test systems for evaluating the shadow MD framework.

More general nonlinear energy functionals, higher-order multipolar interactions, and more challenging systems involving chemical reactions will not be considered in the present numerical demonstrations. Such applications are beyond the scope of this study. Likewise, we will not investigate hardware scaling or conduct detailed performance evaluations of our software implementations.

\subsection{Fourth-order Flexible-Charge Equilibration Model}

To evaluate and demonstrate shadow MD for flexible-charge equilibration models, we first consider a system of liquid water (300 atoms) in a periodic box using the fourth-order charge-equilibration model of Eq.\ (\ref{Eq:4th_order_QEQ}), i.e.\ with $M=4$, and monopole net partial occupations, ${\bf q} \equiv \{q_i\}$. Only the $N$ monopoles without any higher-order multipoles are included. The atomic electronegativities, ${\chi_i}$, were chosen as constants for each atomic species ($\chi_H=0.13$ eV for Hydrogen and $\chi_O=-0.85$ eV for Oxygen ). Because ${\bf q}$ denotes net partial electron occupations, their signs are opposite to those in the standard partial-charge convention. Atom-type-specific values were also used for the higher-order coefficients, $\{u_i^k\}$. Specifically, we chose $u_H^{(2)}=12.055$ eV, $u_O^{(2)}=11.876$ eV, $u_H^{(3)}=12.055/2$ eV, $u_O^{(3)}=11.876/2$ eV, $u_H^{(4)}=12.055/4$ eV, and $u_O^{(4)}=11.876/4$ eV. The third- and fourth-order coefficients were chosen simply as fractions, $1/2$ and $1/4$, respectively, of the corresponding second-order coefficients. The purpose of including higher-order terms in the present study is to demonstrate the behavior of shadow MD for a nonlinear charge-equilibration model, rather than to improve the physical accuracy of the underlying potential. Nevertheless, previous studies have shown that including the third- and fourth-order terms can significantly improve the accuracy over conventional second-order QEq models. \cite{MZhang2009,OdaTakahashi2010}

The charge-independent potential, $V({\bf R})$, was taken from a charge-independent tight-binding model, but could in principle be provided by any suitable classical force field or short-range MLIP. For the Coulomb interaction, we assumed a $1/R$ interaction at long range that is screened at short distances, corresponding to the electrostatic interaction between overlapping Slater-shaped atomic net partial charge densities, with the on-site Coulomb interaction given by $u^{(2)}$. The electronegativities and the second-order coefficients, $u_H^{(2)}$ and $u_O^{(2)}$, were adapted from self-consistent-charge density-functional tight-binding theory, together with the charge-independent tight-binding parameterization available in the LATTE electronic-structure code. \cite{LATTE,AKrishnapriyan17} 

Instead of using the projected preconditioned conjugate-gradient solver in Alg.\ \ref{Alg:PCG}, we used the simplified fixed preconditioning approximation in Eq.\ (\ref{eq:FixedPrecond}) to determine the charge acceleration, ${\bf \ddot n}(t)$. As the preconditioner, ${\bf M}$, we used the regularized form
\begin{align}
{\bf M} &= {\bf G}
-\frac{{\bf G}{\bf 1}{\bf 1}^{\rm T}{\bf G}}
{{\bf 1}^{\rm T}{\bf G}{\bf 1}},\\
{\bf G} &= \big({\bf C}+E_S''({\bf n})+\delta{\bf I}\big)^{-1},
\end{align}
with a relatively large regularization parameter, $\delta = 2$. The preconditioner was updated only every 100 fs of simulation time, corresponding to approximately 10 complete oscillations of the O-H stretching modes in each water molecule. The large regularization, $\delta = 2$, together with the infrequent updates, was deliberately chosen to demonstrate that the charge acceleration, ${\bf \ddot n}(t)$, in the extended harmonic-oscillator dynamics does not need to be determined with high accuracy. However, with less frequent updates of the preconditioner, we started to see a small systematic drift in the total energy.

With the fixed-preconditioning approximation of Eq.\ (\ref{eq:FixedPrecond}), only two Ewald summations are required per MD time step, i.e.\ one for the Coulomb potential and one for the force evaluation. This substantially reduces the computational overhead associated with the self-consistent charge equilibration normally required for higher-order nonlinear models. In fact, the resulting computational cost would then be comparable to that of a model with fixed atomic charges.

\begin{figure}
\centering
        \includegraphics[width=1\columnwidth]{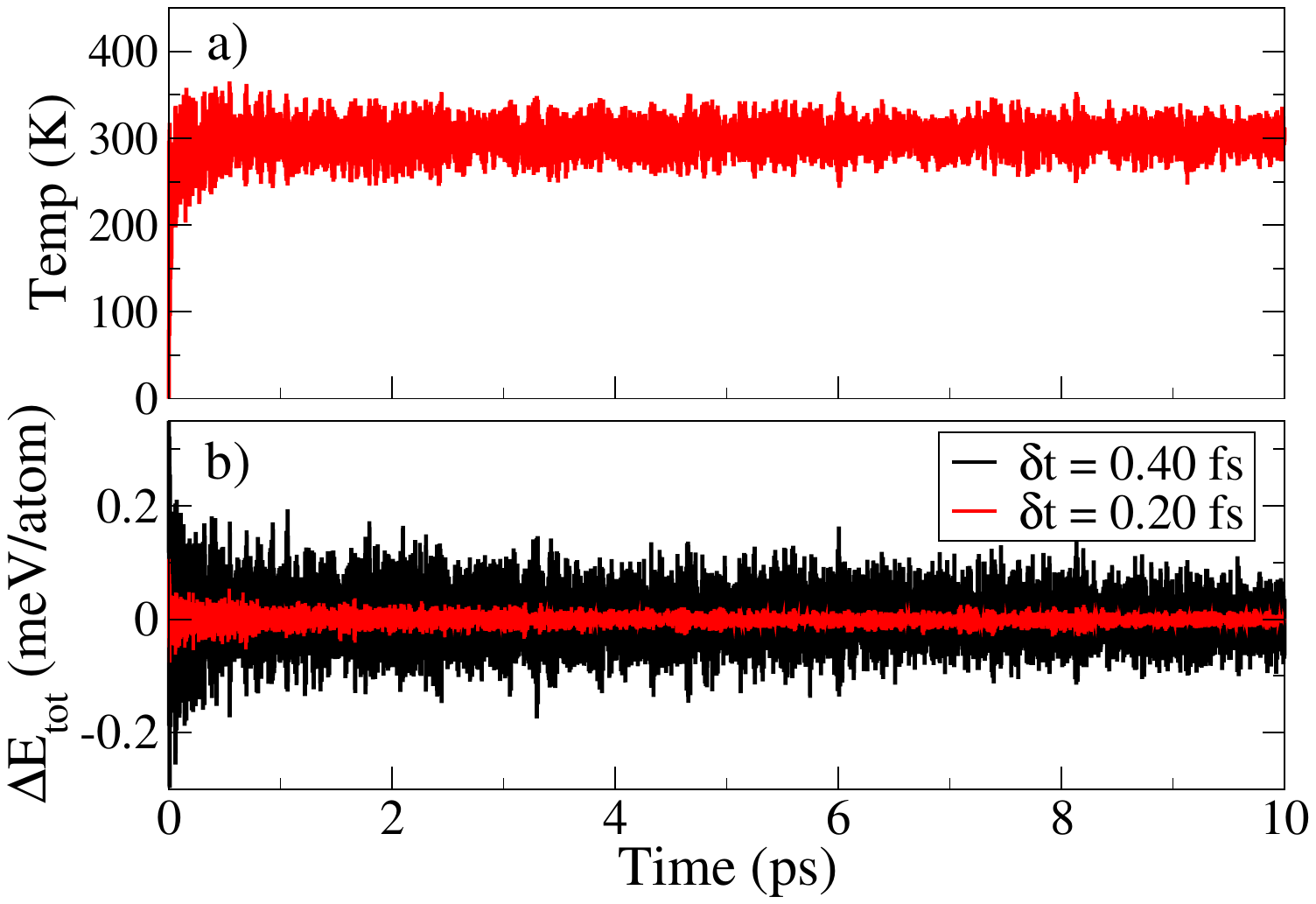}
    \caption{Microcanonical (NVE) shadow MD simulations over 10 ps of liquid water (300 atoms) in a periodic box using the fourth-order monopole charge-equilibration model in Sec.\ \ref{Mth_Order_QEq_Model}. A regularized fixed preconditioner, ${\bf M}$, updated every 100 fs, was used to approximate ${\bf \ddot n}(t)$ according to Eq.\ (\ref{eq:FixedPrecond}). The upper panel a) shows the statistical temperature (red curve), and the lower panel b) shows the fluctuations in the total energy (kinetic + potential) using an integration time step of $\delta t = 0.4$ fs (black curve). For comparison, the total-energy fluctuations obtained with the smaller time step $\delta t = 0.2$ fs are also shown (red curve). }
    \label{fig:100Water}
\end{figure}

\begin{figure}
\centering
    \includegraphics[width=1\columnwidth]{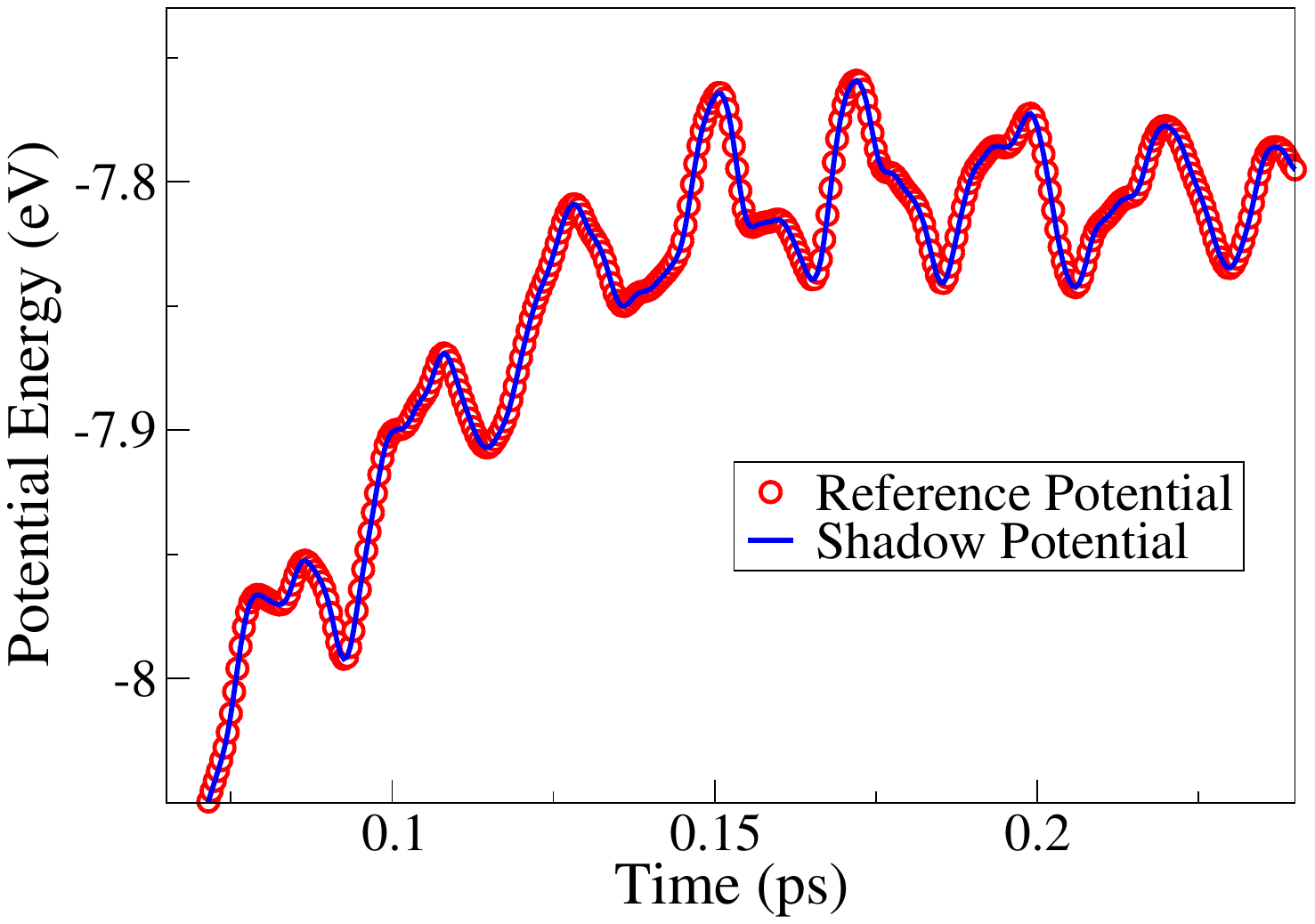}
    \caption{Snapshot of a microcanonical (NVE) simulation of liquid water (300 atoms) in a periodic box using the fourth-order charge-equilibration model in Sec.\ \ref{Mth_Order_QEq_Model}. A regularized fixed preconditioner, ${\bf M}$, updated every 100 fs, was used to approximate ${\bf \ddot n}(t)$ according to Eq.\ (\ref{eq:FixedPrecond}). The charge-dependent part of the potential energy is shown for the tightly converged Born-Oppenheimer reference solution (red circles) and the corresponding shadow potential (blue solid line) along the shadow MD trajectory generated using a time step of $\delta t=0.4$ fs, as in Fig.\ \ref{fig:100Water}. }
    \label{fig:100Water_Pot}
\end{figure}

The shadow MD simulation results are shown in Figs.\ \ref{fig:100Water} and \ref{fig:100Water_Pot}. The upper panel a) of Fig.\ \ref{fig:100Water} shows the statistical temperature (red) determined from the instantaneous kinetic energy, and the lower panel b) shows the fluctuations in the total energy (black). The simulation used an integration time step of $\delta t = 0.4$ fs and a water density of approximately 1 g/cm$^3$.

The total energy remains stable without any visible systematic drift. For comparison, we also show the total-energy fluctuations obtained using $\delta t = 0.2$ fs (red curve). Reducing the time step by a factor of two reduces the magnitude of the energy fluctuations by approximately a factor of four, as expected from the Verlet integration scheme.

Figure \ref{fig:100Water_Pot} shows the charge-dependent part of the shadow potential energy along a trajectory generated by shadow MD using an integration time step of $\delta t = 0.4$ fs with a fixed preconditioning approximation, Eq.\ (\ref{eq:FixedPrecond}). For comparison, we show the corresponding tightly converged Born-Oppenheimer reference potential energy evaluated at each time step along the same shadow MD trajectory using an iterative Newton scheme as the nonlinear solver. The two curves are virtually indistinguishable, demonstrating that the shadow potential closely tracks the corresponding tightly converged Born-Oppenheimer potential-energy surface.

These results demonstrate that only modest accuracy is required when integrating the extended harmonic-oscillator equations, significantly reducing the computational cost. This example also demonstrates how higher-order non-linear models can be incorporated into a MLIP with a cost approaching that of a simple fixed charge model.

In the present example, the regularized preconditioner, ${\bf M}$, is constructed from an exact direct inversion of a fairly large approximate Hessian. For larger systems containing tens of thousands of atoms, however, such a direct solution is no longer practical. Scalable simulations therefore require alternative preconditioning approaches, such as the sparse approximate-inverse or related methods. \cite{Kolotilina1993,Grote97,Benzi96,ANakano97,Benzi2002}

\subsection{Second-order Flexible-Charge Equilibration Model}

We next study the regular second-order flexible-charge equilibration (QEq) model discussed in Sec.\ \ref{QEq_model} with the 2nd-order energy expression in Eq.\ (\ref{Eq:2nd_Order_QEQ}) for liquid water using the parameterization of the model as described by Kaymak \textit{et al.}~\cite{Kaymak2025GPU} 
The fixed atom-specific electronegativity and chemical hardness parameters in this QEq model were fitted to molecular dipole moments \cite{Kaymak2025GPU} and the charge-independent potential, $V({\bf R})$, is represented by a Hierarchically Interacting Particle Neural Network (HIP-NN), \cite{NLubbers18} trained on the difference between the reference energies and forces and the corresponding contributions from the equilibrated QEq model. \cite{Kaymak2025GPU} The long-range Coulomb interactions are evaluated with particle-mesh Ewald (PME) summations under periodic boundary conditions including fast neighborlist calculations using a recent GPU-accelerated software package implemented in Python by Kaymak {\em et al.} \cite{Kaymak2025GPU} However, the shadow MD presented here should, of course, be applicable also together with other recent Ewald summation and neighborlist methods. \cite{MGubler24,PLoche25}

The integration of the harmonic-oscillator equations of motion in shadow MD for the extended electronic degrees of freedom, ${\bf n}(t)$, requires the evaluation of the charge acceleration, ${\bf \ddot n}(t)$. We can determine this acceleration using the projected preconditioned conjugate-gradient (PPCG) method in Alg.\ \ref{Alg:PCG} for the linear system of equations in Eq.\ (\ref{eq:PCG}) or the  preconditioned GMRES method \cite{saad1986gmres} for the system of linear equations in Eq.\ (\ref{Eq:2nd_Order_QEQ_EOM}).

For both GMRES and PPCG, we first use a diagonal preconditioner based on the on-site diagonal contribution to the Hessian, i.e.\ the chemical hardness parameters in the QEq model. 

Apart from the diagonal Jacobi preconditioner we also use a preconditioner constructed from a real- and reciprocal-space separated Ewald summation of the Coulomb potential using Accurate Long-, Average Mid-, and On-site Short-range (ALAMOS) approximations of the inverse Coulomb operator. This ALAMOS preconditioner treats the short-range interactions explicitly using a real-space representation with a thresholded sparse approximate inverse, the long-range is included using an explicit partial sum over a limited set of k-space vectors, and the mid-range is based on a phase-averaged approximation. Details of the ALAMOS preconditioner are presented elsewhere. \cite{Thomas2026} We will use either only the real-space short-range local part of the ALAMOS preconditioner (Local ALAMOS), corresponding to a sparse approximate inverse preconditioner, \cite{Kolotilina1993,Grote97,Benzi96,ANakano97,Benzi2002} or the full complete ALAMOS preconditioner (Full ALAMOS).

\begin{figure}
    \centering
    \includegraphics[width=1\columnwidth]{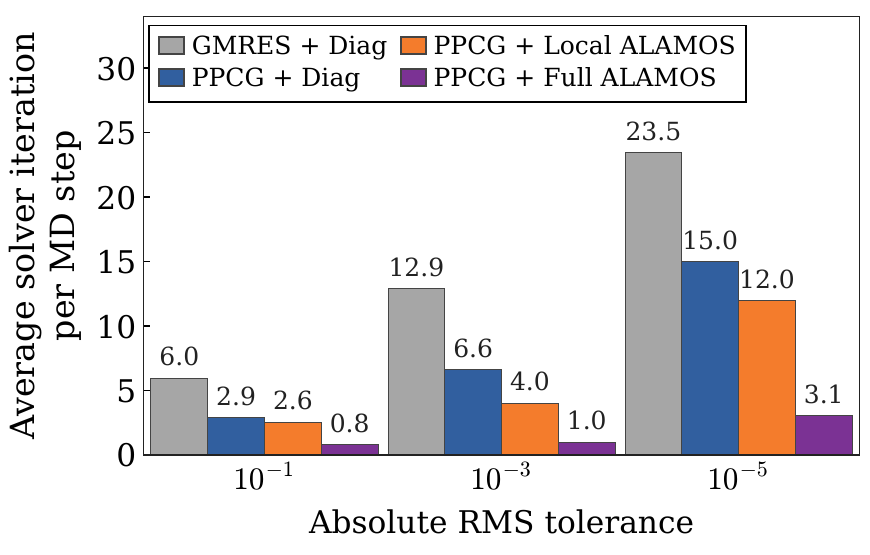}
    \caption{
     Average number of solver iterations per MD step required to calculate ${\bf \ddot n}(t)$ at different convergence tolerances for different solvers and preconditioners for shadow MD simulation with the second-order flexible-charge equilibration (QEq) model. The NVE simulations were performed for a 52,320-atom system using double precision (FP64) with an integration time step of $\delta t=0.5~\mathrm{fs}$. 
    }
    \label{fig:solver_iterations}
\end{figure}

In Fig.\ \ref{fig:solver_iterations}, we show the average number of solver iterations required per MD time step to reach a given convergence tolerance in the calculation of ${\bf \ddot n}(t)$ for a 52,320-atom water system using GMRES with diagonal preconditioning and PPCG with diagonal, local ALAMOS, and full ALAMOS preconditioning. \cite{Thomas2026} The ALAMOS preconditioners were rebuilt every $1000$ MD steps. The first 400 MD steps were excluded as warm-up, and the reported values are averages over the following $2000$ MD steps and three runs.
The results are shown for absolute root-mean-square (RMS) charge-residual convergence tolerances of $10^{-1}$, $10^{-3}$, and $10^{-5}$. With diagonal preconditioning, PPCG requires fewer solver iterations than GMRES over the full range of tolerances. At tolerances of $10^{-1}$ and $10^{-3}$, the average iteration count is reduced by approximately a factor of two. The ALAMOS preconditioners further reduce the number of PPCG iterations, with full ALAMOS requiring on average $0.80$, $1.00$, and $3.06$ iterations at tolerances of $10^{-1}$, $10^{-3}$, and $10^{-5}$, respectively. Although GMRES is computationally more expensive, it is more general and does not require a positive-definite system matrix. This flexibility could be important if the matrix ${\bf C} + {\bf H}_{\rm S}({\bf n})$ in Eq.\ (\ref{Eq:ChargeAcc}) develops negative eigenvalues, as may occur for highly nonlinear energy functionals during complex chemical reactions. However, such cases are beyond the scope of the present study.

\begin{figure}
    \centering
    \includegraphics[width=\columnwidth]{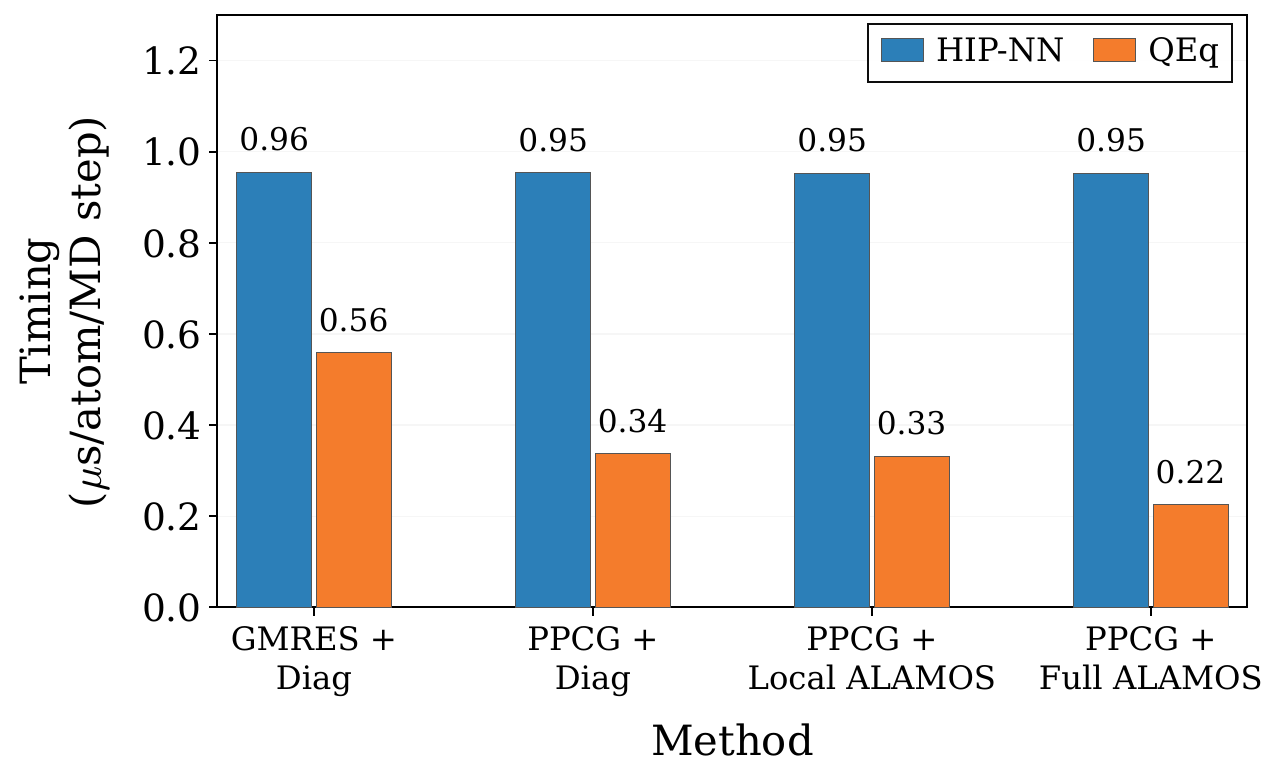}
    \caption{
    Charge-independent HIP-NN and charge-dependent QEq timings for the different methods and preconditioners in Fig.~\ref{fig:solver_iterations} at an absolute RMS charge-residual tolerance of $10^{-1}$. The 52,320-atom NVE simulations were performed in FP64 with an integration time step of $\delta t = 0.5~\mathrm{fs}$ on an NVIDIA A100 GPU. \cite{Kaymak2025GPU} The cost of a Full ALAMOS preconditioner is only about $1~\mu\mathrm{s}/\mathrm{atom}$, which is updated only once every 1000 MD time steps. 
    }
    \label{fig:hipnn_qeq_timing}
\end{figure}

The corresponding timings for calculating the charge-independent potential energies and forces with HIP-NN, as well as the charge-dependent QEq potential energies and forces together with the charge acceleration, ${\bf \ddot n}(t)$, using the different solvers and preconditioners, are shown in Fig. \ref{fig:hipnn_qeq_timing} for the 52,320-atom water system. The convergence criterion for determining the charge accelerations in the shadow MD simulations was set to an RMS charge-residual tolerance of $10^{-1}$. These microcanonical (NVE) simulations were performed in double precision (FP64) with an integration time step of $\delta t = 0.5~\mathrm{fs}$ running on an NVIDIA A100 GPU. The timings were accumulated over $1000$ MD steps following $400$ warm-up steps.

The HIP-NN timing remains approximately $0.95~\mu\mathrm{s}/\mathrm{atom}/\mathrm{MD~step}$ for all four solver and preconditioner combinations. This is a fairly fast charge-independent MLIP running on a single GPU in FP64. \cite{Drautz21, NVIDIA_ALCHEMI_BMD} The wall-clock time for this short-range ML potential could also be significantly reduced, at least in principle, using multiple GPU nodes. However, our timing results for the flexible-charge equilibration models are independent of the choice of the charge-independent MLIP. 

With diagonal preconditioning, PPCG reduces the timing for the charge-dependent parts of the QEq calculations from $0.56$ to $0.34~\mu\mathrm{s}/\mathrm{atom}/\mathrm{MD~step}$ compared to GMRES. The local ALAMOS preconditioner gives a similar QEq timing of $0.33~\mu\mathrm{s}/\mathrm{atom}/\mathrm{MD~step}$, while full ALAMOS further reduces the charge-dependent QEq timing to $0.22~\mu\mathrm{s}/\mathrm{atom}/\mathrm{MD~step}$. The full ALAMOS result is lower by approximately a factor of $2.5$ compared with GMRES using diagonal preconditioning. In this case, the charge-dependent QEq cost is less than one quarter of the HIP-NN cost.

\begin{figure}
    \centering
    \includegraphics[width=\columnwidth]{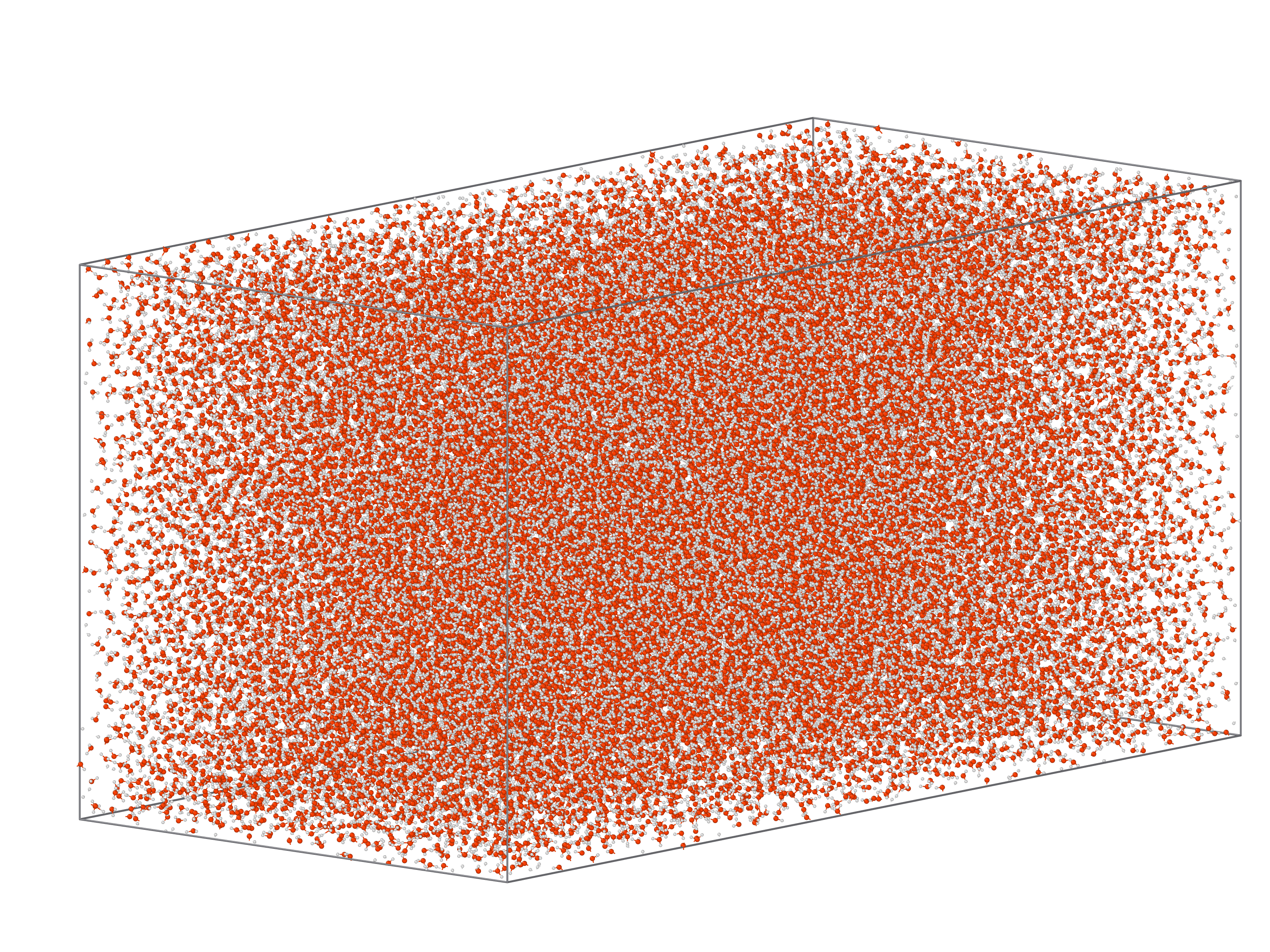}
    \caption{
    Snapshot of the 104,640-atom periodic water system used in the large-scale ALAMOS-0 simulations. Oxygen and hydrogen atoms are shown in red and gray, respectively. The box dimensions are $161.196 \times 80.598 \times 80.598~\mathrm{\AA}^3$ ~ corresponding to a density of about 0.996 g/cm$^3$.
    }
    \label{fig:104k_water}
\end{figure}

We then consider the fixed-preconditioning approximation as in Eq.\ (\ref{eq:FixedPrecond}), denoted ALAMOS-0, in which the full ALAMOS preconditioner is applied directly in a single step without any Krylov iterations. These simulations were performed for a 104,640-atom periodic water system as illustrated in Fig.~\ref{fig:104k_water}. The charge-dependent parts, including the PME, and ALAMOS preconditioner calculations were performed in single precision (FP32), while HIP-NN inference and nuclear propagation were performed in double precision (FP64).

\begin{figure}
    \centering
    \includegraphics[width=\columnwidth]
    {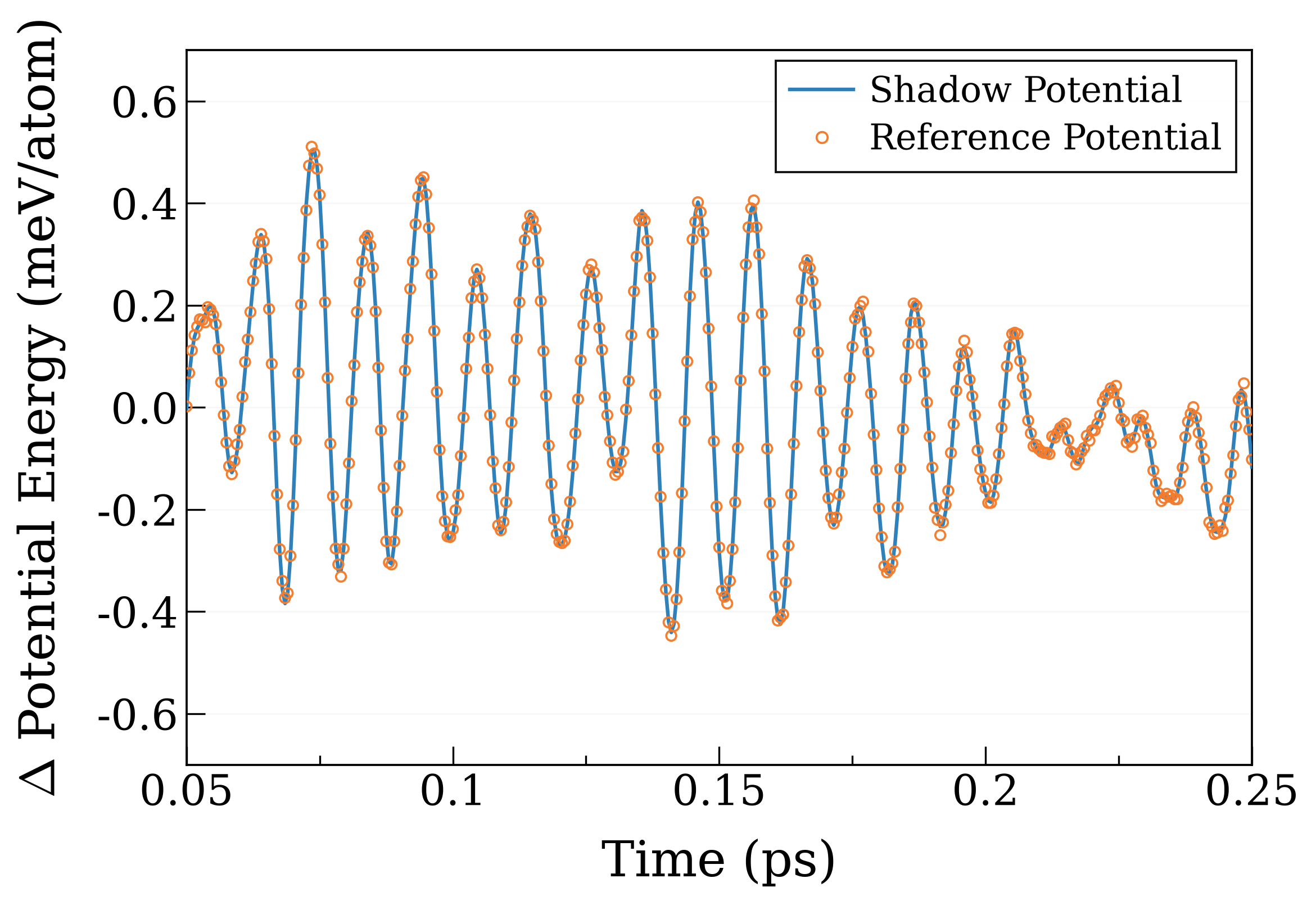}
    \caption{
    Snapshot of the ALAMOS-0 shadow MD simulation in the microcanonical (NVE) ensemble of liquid water containing 104,640 atoms using the fixed preconditioning approximation in Eq.~(\ref{eq:FixedPrecond}). The fluctuations in the charge-dependent potential energy per atom is shown for the ALAMOS-0 shadow potential (blue solid line) and the corresponding regular tightly converged reference potential (orange circles), evaluated at every MD time step using a tightly converged charge solution with a relative tolerance of $10^{-6}$. The integration time step was $\delta t=0.5~\mathrm{fs}$.
    }
    \label{fig:shadow_vs_tight_qeq}
\end{figure}

To assess the accuracy of the shadow potential energy surface in the ALAMOS-0 shadow MD simulation, we first compare its charge-dependent potential-energy fluctuations per atom with the corresponding regular and tightly converged Born-Oppenheimer reference potential energy ('Reference Potential'). Both potential energy fluctuations are evaluated along the same molecular trajectory driven by the shadow MD. The comparison is shown in Fig.~\ref{fig:shadow_vs_tight_qeq} for a short snapshot of the 104,640-atom NVE trajectory. An integration time step of $\delta t=0.5~\mathrm{fs}$ was used, with the full ALAMOS preconditioner rebuilt every 5000 MD steps and without any additional Krylov iterations. The charge-dependent parts, including the PME, and ALAMOS calculations were performed in single-precision, FP32, while HIP-NN inference and nuclear propagation were performed in double precision, FP64. The reference potential was evaluated at every MD time step using a tightly converged charge solution with a relative tolerance of $10^{-6}$. The cost of reconstructing the full ALAMOS preconditioner is about $1~\mu\mathrm{s}$ per atom and MD step. With an update every 5000 MD steps, this cost is negligible.

The shadow potential (solid line) and reference potential energy (circles) show very close agreement along the trajectory shown in Fig.~\ref{fig:shadow_vs_tight_qeq}. This agreement demonstrates that the shadow potential-energy accurately follows the fully equilibrated regular Born-Oppenheimer potential-energy surface even when only a fixed preconditioning approximation is used, Eq.\ (\ref{eq:FixedPrecond}). 

\begin{figure}
    \centering
    \includegraphics[width=\columnwidth]{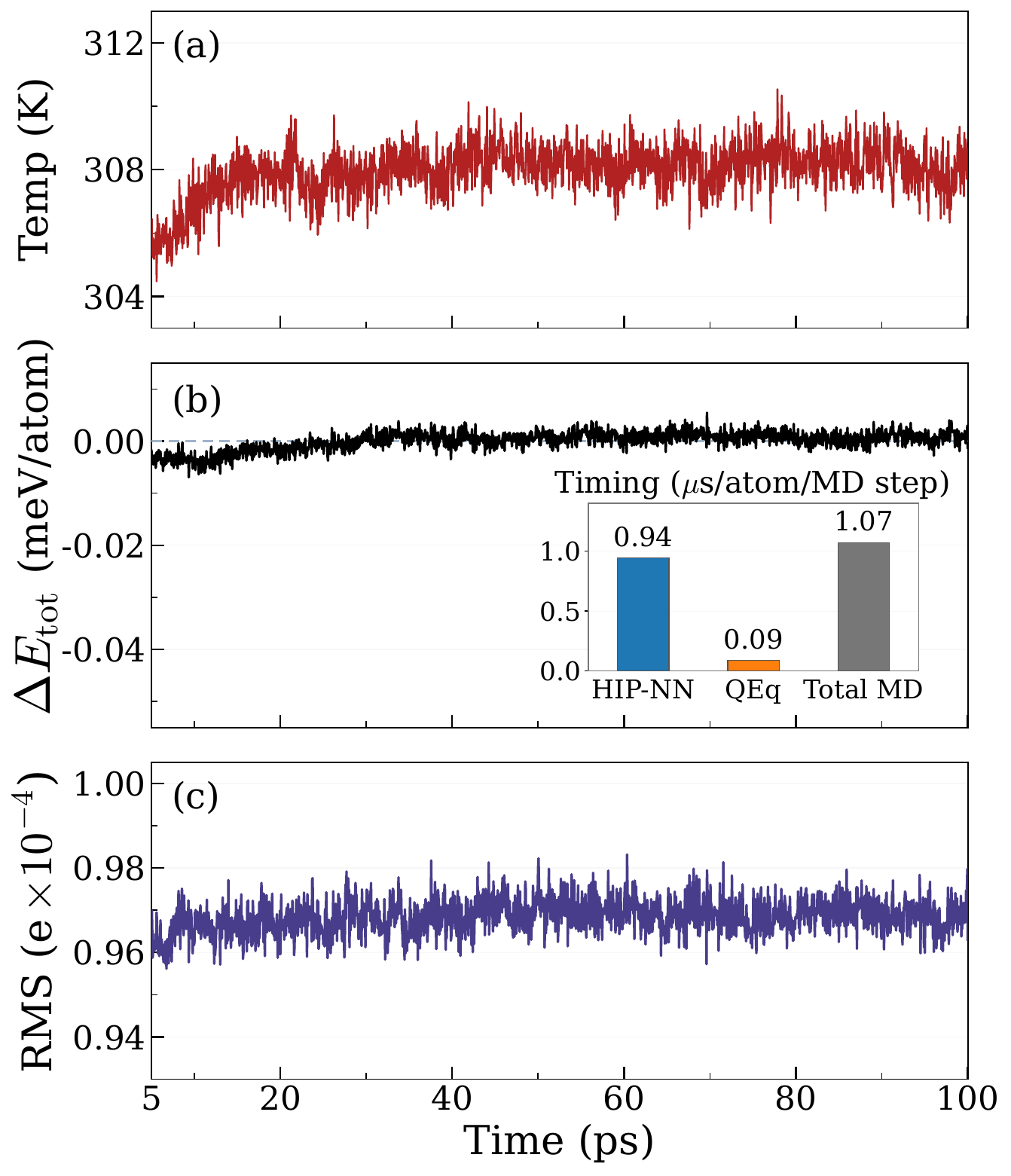}
    \caption{
    A 100 ps microcanonical (NVE) shadow MD simulation with a fixed ALAMOS-0 preconditioning only, Eq.\ (\ref{eq:FixedPrecond}), of a 104,640-atom water system in the microcanonical (NVE) ensemble running on a single Nvidia A100 GPU. The upper panel (a) shows the statistical temperature, the middle panel (b) shows the fluctuations in the total energy per atom, $\Delta E_{\rm tot} = [E_{\rm tot}(t)-E_{\rm tot}(0)]/N$, shifted to have zero mean over the displayed interval, and the lower panel (c) shows the root-mean-square (RMS  in units of $e \times 10^{-4}$) of the charge residual function, $({\bf q}[{\bf n}]-{\bf n})$. 
    The integration time step was $\delta t=0.5~\mathrm{fs}$, with the fixed ALAMOS preconditioner rebuilt every $5000$ MD steps (or 2.5 ps). The cost of constructing the ALAMOS preconditioner is only about $1~\mu\mathrm{s}/\mathrm{atom}$ and is thus negligible over the full MD simulation.  HIP-NN and nuclear propagation were performed in FP64 and the charge-dependent PME and ALAMOS parts of the shadow QEq calculations were performed in FP32. The inset in the middle panel shows separate timing measurements for HIP-NN, QEq, and Total MD in $\mu\mathrm{s}/\mathrm{atom}/\mathrm{MD~step}$.
    }
    \label{fig:alamos0_stability}
\end{figure}

We next examine the long-time stability and computational cost of microcanonical (NVE) shadow MD simulations, using the same fixed ALAMOS-0 preconditioning approximation as in Eq.\ (\ref{eq:FixedPrecond}), for the same 104,640-atom water system shown in Fig.\ \ref{fig:104k_water}.  Figure~\ref{fig:alamos0_stability} shows the statistical temperature, the total energy fluctuations (kinetic + potential) per atom (shifted to have zero mean over the displayed interval), and the root-mean-square charge residual, $\mathrm{RMS}({\bf q}[{\bf n}]-{\bf n})$, over a 100 ps microcanonical NVE simulation. An integration time step of $\delta t = 0.5~\mathrm{fs}$ was used, with the ALAMOS preconditioner rebuilt every $5000$ MD steps and no Krylov iterations were used. The first 5 ps are omitted to exclude the initial transient. HIP-NN inference and nuclear propagation were performed in double precision (FP64) and the charge-dependent PME and ALAMOS parts of the shadow QEq calculations were performed in single precision (FP32). The inset shows the corresponding timings for HIP-NN, the charge-dependent shadow QEq calculations including its force evaluations, and the total shadow MD timings per atom and time step (Total MD).

Following the initial partial equilibration, the statistical temperature fluctuates around $308~\mathrm{K}$, while the total energy fluctuations remain small and bounded, with no evident systematic long-term drift. The RMS charge residual of the residual ${\bf q}[{\bf n}]-{\bf n}$ remains close to $1\times10^{-4}~e$ and shows no systematic growth or instabilities. The average timing results give $0.94~\mu\mathrm{s}/\mathrm{atom}/\mathrm{MD~step}$ for HIP-NN, $0.09~\mu\mathrm{s}/\mathrm{atom}/\mathrm{MD~step}$ for the total charge-dependent shadow QEq calculations, and $1.07~\mu\mathrm{s}/\mathrm{atom}/\mathrm{MD~step}$ for the complete MD step. Thus, the contribution from the 2nd-order shadow QEq model represents less than $10\%$ of the total computational cost. Together with the close agreement between the shadow and tightly converged reference potentials shown above, these results demonstrate both highly accurate and efficient shadow MD simulations with excellent long-term stability.

Generating a stable shadow MD for water systems using only a fixed preconditioning approximation, Eq.\ (\ref{eq:FixedPrecond}), was not possible in previous studies for flexible charge models, apart from dipole-only shadow MD simulations by Grove {\em et al.} \cite{RGrove26} In these previous studies two to three preconditioned Krylov iterations (or low-rank updates) were needed to achieve stable trajectories even for much smaller water systems (with only up to 128 water molecules). \cite{JGoff23, LiNiklasson25, RGrove26} With the new formulation and ALAMOS preconditioner, we find excellent stability and performance, where the cost of including the shadow QEq model to the MLIP is only a small fraction of the total cost per MD time step for a simulation of a system containing 104,640 atoms of water running on a single Nvidia A100 GPU. For larger, distributed computing platforms with multiple GPU nodes this balance may shift. While the short-range charge-independent part of the potential is easy to parallelize, this is not true for the PME calculations. The fast Fourier transform of the PME method is often very hard to parallelize efficiently, but other techniques such as the fast multipole method could be used alternatively. \cite{LGreengard87,Kohnke20} 

The key result of the simulation examples above is that shadow MD is virtually indistinguishable from a tightly converged regular Born-Oppenheimer scheme even when the integration of the extended dynamical charge evolution in Eq.\ (\ref{Eq:ChargeAcc}) is approximate. In this way it is possible to generate stable MD trajectories using only a fixed projected preconditioning approximation as in Eq.\ (\ref{eq:FixedPrecond}). This approach requires only two full Ewald (or Coulomb) summations, one for the potential energy and one for the force evaluations, in each MD time step, which is only half the number of Ewald summations required in a previous study of a small water system using the same parameterization \cite{Kaymak2025GPU}. The computational cost of including a flexible-charge model with long-range interactions and all-to-all charge equilibration is thus on par with a simple fixed charge model.

\subsection{Shadow Langevin Molecular Dynamics}

We lastly demonstrate how shadow MD can be combined with Langevin dynamics to simulate the canonical (NVT) ensemble. We use the Gr\o{}nbech-Jensen GJ-I Langevin integration scheme \cite{NGronbechJensen20,Finkelstein2021GJ} (or, originally known as GJF \cite{NGronbechJensen12}), where the nuclear degrees of freedom are coupled to a heat bath with friction coefficient of $\gamma=1~\mathrm{ps}^{-1}$. This method is part of a unique, broader collection of Verlet-based Langevin integrators, which yield statistics independent of both friction $\gamma$ and time step $\Delta t$ in the free-diffusion and linear case.\cite{NGronbechJensen20,Finkelstein2021GJ} The same second-order flexible-charge equilibration/HIP-NN water model as above by Kaymak {\em et al.} is used for a system containing 900 atoms (300 water molecules) at target temperatures of 100~K and 300~K. \cite{Kaymak2025GPU} The instantaneous kinetic temperature was evaluated as $T_{\mathrm{kin}} = 2K/(3Nk_{\mathrm{B}})$, where $K$ is the nuclear kinetic energy calculated using the GJ-I half-step velocities as described in Ref.~\onlinecite{Finkelstein2021GJ}. No center-of-mass corrections were applied.

The initial states at 300~K and 100~K were prepared using Langevin dynamics with a fixed ALAMOS-0 preconditioning approximation for the shadow MD, Eq.\ (\ref{eq:FixedPrecond}). The system was first propagated for 20~ps at 300~K, and the resulting nuclear positions and velocities were then used to initialize the ALAMOS-0 at 300~K. Starting from this state, the system was propagated for an additional 20~ps at 100~K to prepare the corresponding 100~K state, which was then used to initialize both simulations at 100~K. Thus, the ALAMOS-0 shadow Langevin MD simulations at 100 K and 300 K were initialized from their respective temperature-prepared nuclear positions and velocities.

\begin{figure}
    \centering
    \includegraphics[width=\columnwidth]    {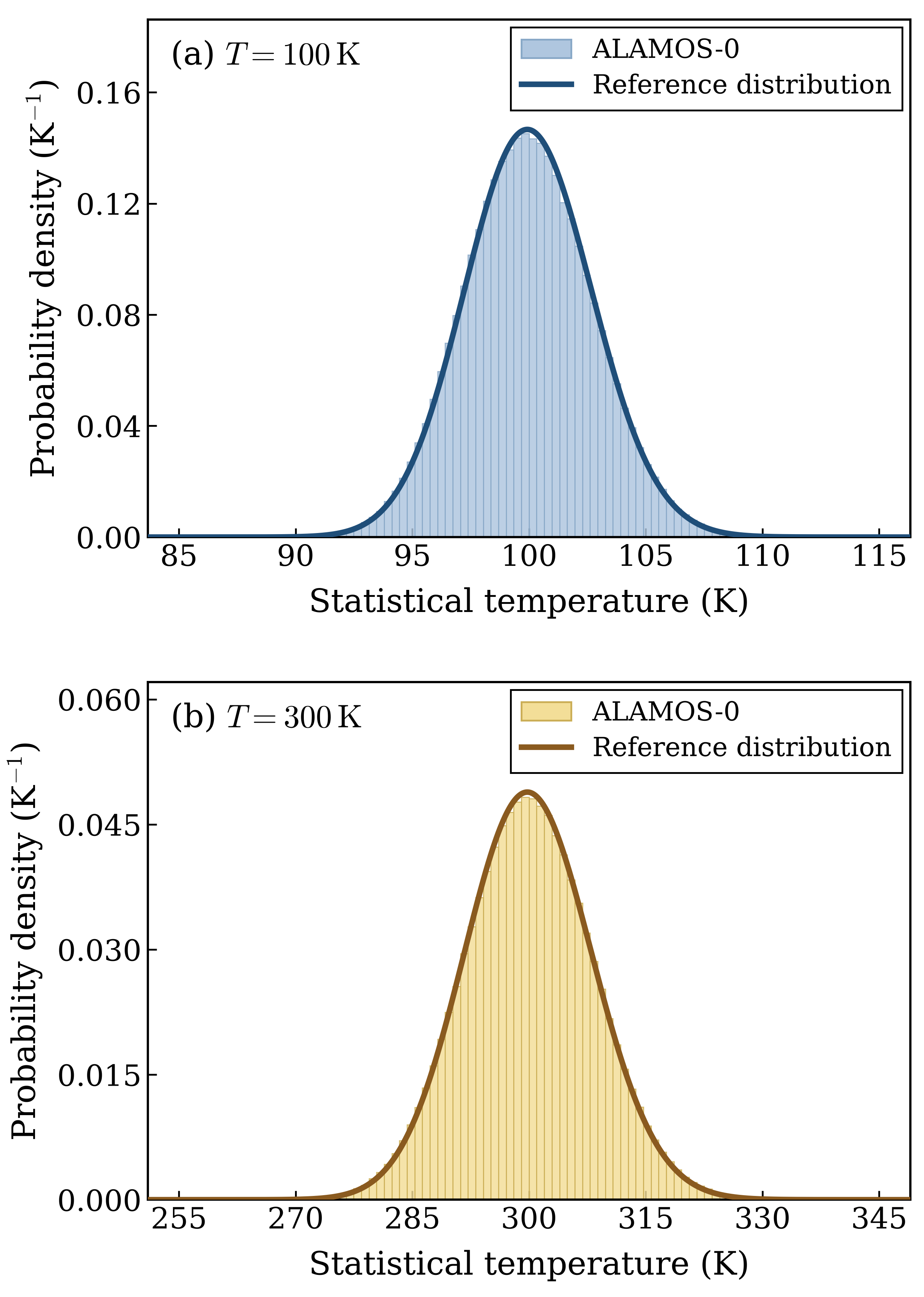
}
    \caption{Statistical distribution of the instantaneous kinetic temperature sampled from shadow Langevin MD (NVT) simulations of 300 water molecules using a fixed preconditioning approximation ALAMOS-0, Eq.\ (\ref{eq:FixedPrecond}), at target temperatures of (a) 100~K and (b) 300~K over a period of 1 ns. The integration time step was $\delta t=0.25$~fs, with the ALAMOS preconditioner rebuilt every $5000$ MD steps (or 1.25 ps). The solid lines correspond to the reference distribution expected from statistical theory. \cite{Merz2018}}
    \label{fig:nvt_temperature_distribution}
\end{figure}

Figure~\ref{fig:nvt_temperature_distribution} shows the sampled distribution of the instantaneous kinetic temperature obtained from shadow Langevin MD simulations using the ALAMOS-0 fixed preconditioning approximation, Eq.\ (\ref{eq:FixedPrecond}). 
The distributions were obtained over 1 ns of simulation time at target temperatures of 100 K (upper panel, a) and 300 K (lower panel, b) after an initial equilibration of 100 ps using an integration time step $\delta t = 0.25$ fs. For each temperature interval, the number of sampled instantaneous temperatures falling within the corresponding bin is counted, and the resulting binned temperature distribution is thereafter normalized.

For a system in the canonical ensemble, the individual momentum components follow Gaussian distributions, and the resulting total kinetic energy follows a Gamma distribution. \cite{Merz2018} Consequently, the instantaneous kinetic temperature is also Gamma distributed, though due to the central limit theorem, becomes normally distributed at large system sizes.
The sampled temperature fluctuations in Fig.\ \ref{fig:nvt_temperature_distribution} are closely aligned with the expected theoretical reference values shown by the solid lines and are in close agreement with Gaussian distributions centered around the target temperatures, $T_0$, with standard deviations, $\sigma = T_0\sqrt{2/(3N)}$, where $N$ is the number of atoms. No significant deviations or instabilities appear because of the shadow MD formulation in combination with the statistical noise present in the Langevin dynamics. Alternative PPCG-based Langevin MD simulations (not shown) gave the same close agreement with the theoretical predictions coming from the canonical partition function. These findings are all consistent with related quantum-based extended Lagrangian Born-Oppenheimer MD simulations using both stochastic and deterministic thermostats. \cite{EMartinez15}

\begin{figure}
    \centering
    \includegraphics[width=\columnwidth]{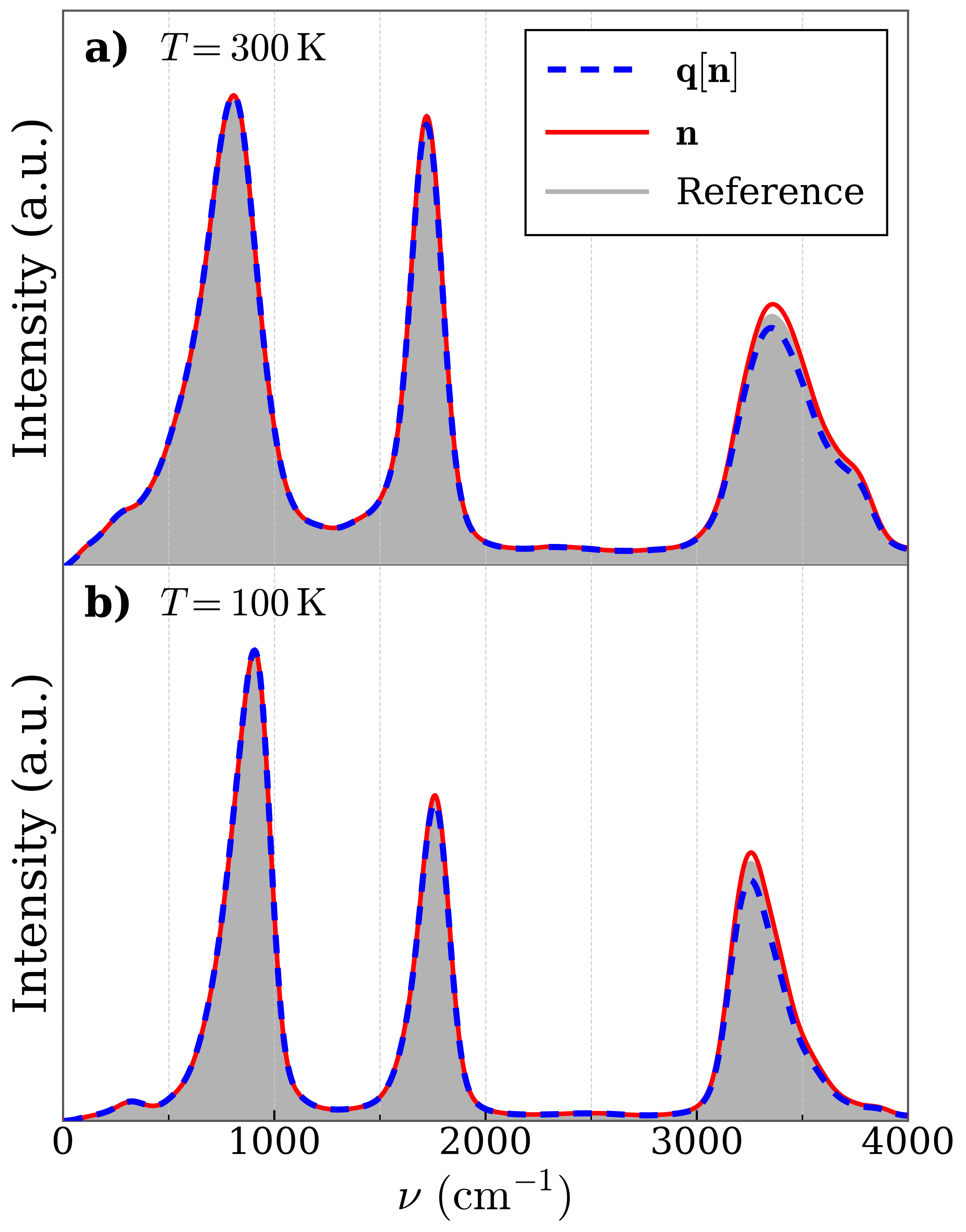}
    \caption{IR spectra from 100 ps NVT simulations using a Langevin shadow MD for 300 water molecules at approximately 300 K in (a) and 100 K in (b). The IR spectra are calculated from the dipole autocorrelation function using the  dynamical charges  ${\bf n}$, the corresponding optimized shadow-potential charges ${\bf q}[{\bf n}]$, and tightly converged reference charges (Reference). No quantum corrections were included. The trajectories were generated with a time step
    of $\delta t=0.5$ fs using a GJ-I Langevin thermostat with $\gamma=1~\mathrm{ps}^{-1}$.}
    \label{fig:ir_spectra}
\end{figure}

To further evaluate the ability of the shadow MD to capture fluctuating charges without artifacts within an NVT ensemble based on a stochastic Langevin dynamics we also calculate the infrared (IR) spectra.  
For the IR analysis, we use 100~ps of equilibrated ALAMOS-0 trajectories at 100~K and 300~K. The total dipole moment was evaluated using the propagated dynamical charges ${\bf n}$, the corresponding optimized shadow-potential charges ${\bf q}[{\bf n}]$, and tightly converged regular QEq reference charges evaluated at the same nuclear configurations. The regular QEq Born-Oppenheimer reference charges were converged to a relative tolerance of $10^{-10}$. The spectra were obtained from the Fourier transform of the total dipole-moment autocorrelation function, following established time-correlation-function approaches for calculating the IR spectra from MD trajectories. \cite{Mukamel1995,Baiz2020,Gaigeot2003,Gaigeot2007} 
The dipole autocorrelation function was estimated by dividing each 100 ps trajectory into 200 consecutive, non-overlapping windows of 0.5 ps each, containing 1000 samples per window. The autocorrelation functions were calculated separately for each window and then averaged before Fourier transformation to obtain the IR spectrum. No quantum corrections were applied. Figure~\ref{fig:ir_spectra} shows the resulting spectra at approximately 300~K and 100~K.

The spectra obtained from ${\bf n}$ and ${\bf q}[{\bf n}]$ closely reproduce the reference results at both temperatures. The characteristic low-frequency, H--O--H bending ($900~\mathrm{cm}^{-1}$ region), and O--H stretching ($1700~\mathrm{cm}^{-1}$ region) features occur at the same wavenumbers for all three charge representations. This agreement shows that the extended dynamical charge variables of the shadow MD reproduce the characteristic vibrational frequencies of the tightly converged QEq reference. No significant frequency shifts or additional peaks appear from the dynamical charges in the shadow MD. However, we notice some smaller differences in the amplitude for the high frequency peaks corresponding to intramolecular O--H stretches above $3000$ cm$^{-1}$. At these frequencies we start to see small errors due to the finite size or the integration time step. A shorter integration time step could be used to reduce the differences.

\begin{figure}
    \centering
    \includegraphics[width=\columnwidth]{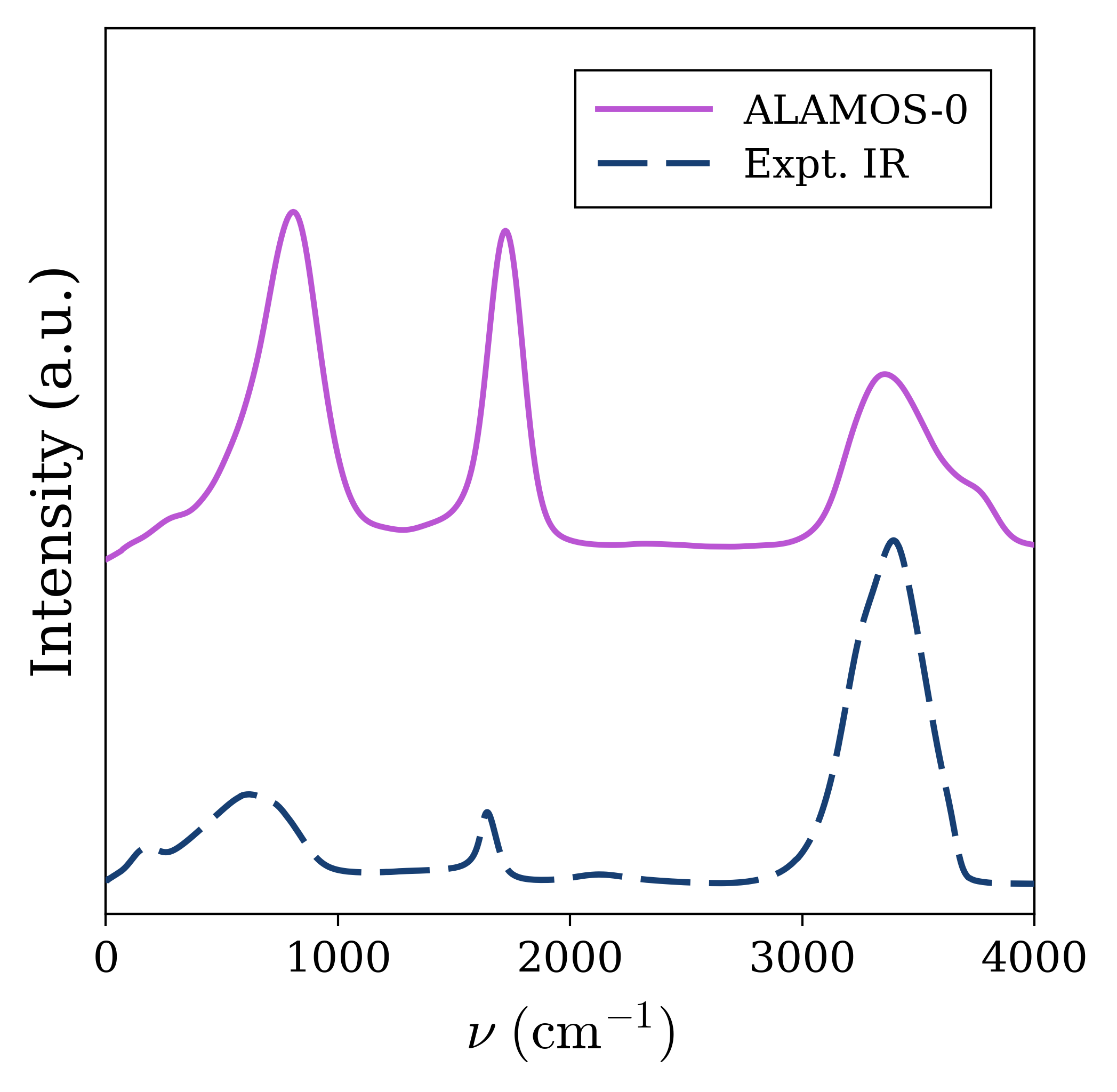}
    \caption{The water IR spectrum obtained from the 300~K ALAMOS-0 trajectory of the Langevin shadow MD simulation in comparison with the experimental spectrum of liquid water at $25^\circ$C. \cite{Bertie1996} No quantum corrections were included. The spectra are normalized and vertically displaced for clarity. Improved agreement can be achieved using environment-dependent atomic electronegativities and chemical hardness parameters. \cite{LiNiklasson25}}
    \label{fig:ir_alamos_exp}
\end{figure}

Finally, we compare the IR spectrum obtained from the 300~K ALAMOS-0 trajectory of the Langevin shadow MD simulation with the experimental infrared spectrum of liquid water reported by Bertie and Lan. \cite{Bertie1996} The calculated spectrum uses the tightly converged Born-Oppenheimer QEq reference charges evaluated along the ALAMOS-0 trajectory and the same 200-window analysis described above with an integration time step, $\delta t = 0.5~$fs. The comparison is shown in Fig.~\ref{fig:ir_alamos_exp}. No quantum corrections were applied.

The calculated spectrum reproduces the characteristic low-frequency intermolecular, H--O--H bending, and O--H stretching regions observed experimentally. The corresponding features occur at similar wavenumbers, with particularly close agreement for the O--H stretching band near $3400~\mathrm{cm}^{-1}$. Thus, the calculated spectrum reproduces the characteristic vibrational frequencies of liquid water demonstrating a realistic and fairly accurate water simulation based on a MLIP combining the HIP-NN potential and the 2nd-order flexible-charge equilibration model. Similar agreements with experiments are also found for the water model in comparison with the radial distribution functions. \cite{Kaymak2025GPU} These simulations demonstrate a realistic and fairly accurate MLIP for water that also includes long-range charge-dependent response. To get a more accurate representation we would need atomic electronegativities and chemical hardness parameters that depend on their local environments. \cite{LiNiklasson25}

\section{Shadow Molecular Dynamics from a Backward-Error Perspective}
\label{sec:Perspective}

Backward error analysis provides a natural perspective for understanding shadow MD. \cite{Wilkinson1963,SReich99,golub2013matrix} In geometric numerical integration of Hamiltonian dynamics, the numerical trajectories that are generated by a symplectic integrator can be viewed as the exact trajectories of a nearby \emph{shadow} Hamiltonian rather than as approximate trajectories of the original Hamiltonian. \cite{HYoshida90,CGrebogi90,BenettinGiorgilli1994,SToxvaerd94,JGans00,BJLeimkuhler04,RDEngel05,HairerLubichWanner2006,SToxvaerd12,BLeimkuhler15} The existence of this nearby Hamiltonian explains the excellent long-time energy conservation and stability of symplectic or geometric integration schemes.

Our formulation of shadow MD adopts the same underlying idea, but is based on a fundamentally different form of approximation. Instead of introducing a shadow system indirectly through the discretized time-integration of Hamilton's equations, the approximation is introduced directly into the nonlinear electronic variational problem that defines the Born-Oppenheimer potential energy surface. The exact nonlinear electronic energy functional is replaced by a nearby shadow functional obtained through a partial linearization, and the exact charge equilibration of this shadow functional defines a nearby shadow Born-Oppenheimer potential-energy surface.

This distinction is important. In geometric or symplectic integration schemes the modified Hamiltonian is an implicit consequence of the numerical integration and depends explicitly on the integration time step. In contrast, the shadow Born-Oppenheimer surface introduced here is constructed for a non-linear flexible-charge equilibration problem and is independent of the choice of the molecular dynamics integration scheme. The shadow potential therefore exists regardless of how the equations of motion are propagated. Any suitable time-reversible geometric integration scheme can thus be used to evolve the dynamics on the shadow potential. \cite{ANiklasson08,AOdell11,AAlbaugh18}

The accuracy of the shadow approximation is controlled by extended electronic dynamical variables introduced through the extended-Lagrangian formulation. These extended dynamical variables, ${\bf n}(t)$, evolve together with the nuclei and continuously update the linearization point of the electronic energy functional, allowing the shadow potential to closely follow the exact Born-Oppenheimer potential-energy surface throughout the simulation. As the extended variables approach the exact electronic ground state, the shadow potential converges systematically toward the exact Born-Oppenheimer potential. \cite{ANiklasson21b}


The shadow MD framework presented here can be further extended using, for example, first-level updated shadow Born-Oppenheimer potentials, \cite{Niklasson23,JGoff23} which further reduce the error relative to the exact Born-Oppenheimer potential. However, these higher-level shadow potentials require at least one additional Coulomb summation and are not considered in the present study.

\section{Summary and Conclusions}
\label{sec:Summary}

Modern MLIPs have expanded the reach of atomistic MD simulations by combining high accuracy with computational efficiency, enabling accelerated discovery across a broad range of problems in materials science, chemistry and molecular biology. Despite this progress, an efficient and transferable description of long-range charge transfer, polarization, and electrostatic interactions remains challenging. These effects are essential in many complex chemical environments, including electrochemical interfaces, ionic materials, heterogeneous catalysts, molecular crystals with defects, and biological systems. In such environments, the electronic charge distribution is governed not only by the local atomic environments but also by a global electronegativity equilibrium involving all-to-all Coulomb interactions. Physics-motivated flexible-charge equilibration models based on DFT provide a natural general framework for capturing this nonlocal, nonlinear electronic response, but the required global equilibration can be computationally demanding and can destabilize MD simulations unless solved to high accuracy at every time step.

In this article, we have presented a shadow MD framework, governed by the equations of motion in Eqs.\ (\ref{Eq:d2Rdt2}) and (\ref{Eq:ChargeAcc}), or equivalently Eqs.\ (\ref{Eq:d2Rdt2_b}) and (\ref{Eq:d2ndt2_b}), that overcomes these computational challenges for general nonlinear flexible-charge models incorporating long-range multipolar electrostatic interactions and arbitrary local nonlinearities. The central idea is to replace the exact Born-Oppenheimer potential energy surface with a nearby shadow potential obtained by exactly equilibrating a partially linearized electronic energy functional. This construction provides conservative interatomic forces and replaces the nonlinear self-consistent charge-equilibration problem with a nearby quasi-diagonal linear problem that can be solved directly at low computational cost. Instead, auxiliary electronic degrees of freedom are propagated through extended-Lagrangian dynamics, which may require iterative linear solutions to determine their accelerations. However, these solutions do not need to be tightly converged and can often be approximated in a single step using a fixed preconditioner.

Beyond its greater generality, e.g.\ as described by Eqs.\ (\ref{Eq:d2Rdt2_b}) and (\ref{Eq:d2ndt2_b}), which incorporates a broad class of electronic energy functionals, the present formulation improves upon previous shadow MD approaches in computational efficiency, robustness to approximations, and ease of implementation. Rather than relying on specialized low-rank kernel approximations, it can use standard off-the-shelf Krylov subspace solvers such as GMRES and PPCG. The formulation is more robust to approximate integration of the charge degrees of freedom and, in many of our examples, permits the use of a simple fixed preconditioning approximation. We have also shown how the shadow MD can be combined with a stochastic Langevin dynamics for canonical (NVT) simulations. Together, these advances broaden the range of practical applications, enable longer simulation times, allow studies of larger and more complex systems with more advanced higher-order energy functionals, while reducing the computational cost and simplifying the implementation.

Our numerical simulations show that the shadow potential energy surfaces closely track the corresponding tightly converged Born-Oppenheimer reference surfaces and maintain excellent long-term energy stability. In several examples, long-range flexible-charge equilibration can be included at a computational cost comparable to that of a simple fixed-charge model, with stable simulations achieved even using single-precision FP32 arithmetic. In some cases, the additional computational cost is only about 10\% of that of relatively fast neural-network MLIPs used for the short-range charge-independent part of the energy and forces. GPU-accelerated simulations of liquid water containing over 100,000 atoms further demonstrate that the shadow MD approach is practical for large-scale atomistic simulations. In these simulations the preconditioner can be constructed at a cost of only about $1~\mu\mathrm{s}/\mathrm{atom}$ and needs to be updated only once every few thousand MD time steps.

The significance of the shadow MD framework is that the global electronic equilibration no longer needs to be a computational bottleneck in large-scale atomistic simulations. Nonlinear charge transfer, polarization, and long-range electrostatics can be included through an exactly equilibrated shadow potential without performing any iterative charge or multipole optimization at each MD step, while the extended electronic dynamics can be propagated approximately at a low computational cost. This removes a major computational limitation to combining modern MLIPs with physics-based descriptions of the nonlocal electronic response. Shadow MD thus provides a practical approach toward a new generation of high-fidelity atomistic models that maintains the speed and scalability of modern MLIPs. This extends the reach of MLIPs beyond local interactions to include the global, nonlocal electronic charge redistribution and response required to accurately capture the physics and chemistry of atomistic systems in complex, extended molecular environments.

\section{Acknowledgements}

This work is supported by the U.S. Department of Energy Office of Basic Energy Sciences (FWP LANLE8AN), the Los Alamos National Laboratory LDRD program, and the U.S. Department of Energy through Los Alamos National Laboratory (LANL). Discussions with Ralf Drautz, Johannes Margraf, Alexandre Tkatchenko, Steven Valone and Susan Atlas are greatly appreciated. LANL is operated by Triad National Security, LLC, for the National Nuclear Security Administration of the U.S. Department of Energy under Contract No.\ 892333218NCA000001.

\section{Author Declarations} 

\subsection{Conflict of Interest}
The authors have no conflicts to disclose.

\subsection{Data Availability}
The Python codes associated with
the manuscript will be made available as part of the open-source SEDACS package (https://github.com/lanl/sedacs). Other data that support the findings are available from the corresponding author upon reasonable request.

\appendix
\section*{Appendix}

\setcounter{subsection}{0}

\renewcommand{\thesubsection}{\Alph{subsection}}

\subsection{The Jacobian in the adiabatic limit}

To derive the Jacobian ${\bf J}$ in \cref{Eq:Adiabatic_J}, we begin with the definition in \cref{Eq:J_Def1,Eq:J_Def2} and the equation that determines ${\bf q}_0[{\bf n}]$ in \cref{Eq:q(n)}, i.e.\
\begin{align}
\begin{bmatrix}
E''_{\rm S}({\bf n}) & {\bf 1}\\
{\bf 1}^{\rm T} & 0
\end{bmatrix}
\begin{bmatrix}
{\bf q}_0[{\bf n}]\\
\mu_0[{\bf n}]
\end{bmatrix}
= -
\begin{bmatrix}
E'_{\rm S}({\bf n}) - E''_{\rm S}({\bf n}){\bf n} + {\bf C}{\bf n} \\
-Q_{\rm tot}
\end{bmatrix}.
\end{align}
Subtracting
\begin{align}
\begin{bmatrix}
E''_{\rm S}({\bf n}) & {\bf 1}\\
{\bf 1}^{\rm T} & 0
\end{bmatrix}
\begin{bmatrix}
{\bf n}\\
\nu
\end{bmatrix}
\end{align}
from both sides gives
\begin{align}
\begin{bmatrix}
E''_{\rm S}({\bf n}) & {\bf 1}\\
{\bf 1}^{\rm T} & 0
\end{bmatrix}
\begin{bmatrix}
{\bf q}_0[{\bf n}]-{\bf n}\\
\mu_0[{\bf n}]-\nu
\end{bmatrix}
= -
\begin{bmatrix}
E'_S({\bf n}) + {\bf C}{\bf n} + \nu{\bf 1} \\
{\bf 1}^{\rm T}{\bf n} - Q_{\rm tot}
\end{bmatrix}.
\label{Eq:Delta_x_relation}
\end{align}
Taking a variation of both sides gives 
\begin{align}
& \delta
\begin{bmatrix}
E''_{\rm S}({\bf n}) & {\bf 1}\\
{\bf 1}^{\rm T} & 0
\end{bmatrix}
\begin{bmatrix}
{\bf q}_0[{\bf n}]-{\bf n}\\
\mu_0[{\bf n}]-\nu
\end{bmatrix}
+
\begin{bmatrix}
E''_{\rm S}({\bf n}) & {\bf 1}\\
{\bf 1}^{\rm T} & 0
\end{bmatrix}
\delta
\begin{bmatrix}
{\bf q}_0[{\bf n}]-{\bf n}\\
\mu_0[{\bf n}]-\nu
\end{bmatrix} \nonumber \\
&~~~~~~~~~~~~~= - \delta
\begin{bmatrix}
E'_{\rm S}({\bf n}) + {\bf C}{\bf n} + \nu{\bf 1} \\
{\bf 1}^{\rm T}{\bf n} - Q_{\rm tot}
\end{bmatrix}.
\end{align}
In the adiabatic limit (or simply assuming a fixed point approximation), 
\begin{equation}
\left|
\begin{bmatrix}
{\bf q}_0[{\bf n}]-{\bf n}\\
\mu_0[{\bf n}]-\nu
\end{bmatrix}
\right|
\rightarrow
\begin{bmatrix}
{\bf 0}\\
0
\end{bmatrix}.
\end{equation}
Using the definition of the Jacobian,
\begin{equation}
{\bf J}
\begin{bmatrix}
\delta{\bf n}\\
\delta\nu
\end{bmatrix}
=
\delta
\begin{bmatrix}
{\bf q}_0[{\bf n}]-{\bf n}\\
\mu_0[{\bf n}]-\nu
\end{bmatrix},
\end{equation}
we then obtain
\begin{align}
\begin{bmatrix}
E''_{\rm S}({\bf n}) & {\bf 1}\\
{\bf 1}^{\rm T} & 0
\end{bmatrix}
{\bf J}
\begin{bmatrix}
\delta{\bf n}\\
\delta\nu
\end{bmatrix}
=
-
\begin{bmatrix}
{\bf H}_{\rm S}({\bf n})+{\bf C} & {\bf 1}\\
{\bf 1}^{\rm T} & 0
\end{bmatrix}
\begin{bmatrix}
\delta{\bf n}\\
\delta\nu
\end{bmatrix}.
\end{align}
Therefore,
\begin{equation}
{\bf J}
=
-
\begin{bmatrix}
E''_S({\bf n}) & {\bf 1}\\
{\bf 1}^{\rm T} & 0
\end{bmatrix}^{-1}
\begin{bmatrix}
{\bf H}_{\rm S}({\bf n})+{\bf C} & {\bf 1}\\
{\bf 1}^{\rm T} & 0
\end{bmatrix},
\end{equation}
which is the expression for the Jacobian in the adiabatic limit given in \cref{Eq:Adiabatic_J}.

\subsection{Derivation of the Equations of Motion}

To derive the harmonic-oscillator equations of motion in \cref{Eq:ChargeAcc} for the charge degrees of freedom, we insert ${\bf J}$ from \cref{Eq:Adiabatic_J} into the equations of motion in \cref{Eq:AdiabaticEOM}. This gives
\begin{align}
\begin{bmatrix}
{\bf C}+{\bf H}_{\rm S}({\bf n}) & {\bf 1}\\
{\bf 1}^{\rm T} & 0
\end{bmatrix}
\begin{bmatrix}
{\bf \ddot n}\\
{\ddot\nu}
\end{bmatrix}
=
\omega^2
\begin{bmatrix}
E''_{\rm S}({\bf n}) & {\bf 1}\\
{\bf 1}^{\rm T} & 0
\end{bmatrix}
\begin{bmatrix}
{\bf q}_0[{\bf n}]-{\bf n}\\
\mu_0[{\bf n}]-\nu
\end{bmatrix}.
\end{align}
Using \cref{Eq:Delta_x_relation} to substitute for the right-hand side, we obtain
\begin{align}
\begin{bmatrix}
{\bf C}+{\bf H}_{\rm S}({\bf n}) & {\bf 1}\\
{\bf 1}^{\rm T} & 0
\end{bmatrix}
\begin{bmatrix}
{\bf \ddot n}\\
{\ddot\nu}
\end{bmatrix}
= - \omega^2
\begin{bmatrix}
E'_{\rm S}({\bf n}) + {\bf C}{\bf n} + \nu{\bf 1}
\\
{\bf 1}^{\rm T}{\bf n} - Q_{\rm tot}
\end{bmatrix},
\end{align}
which is the equation of motion given in \cref{Eq:ChargeAcc}.

\bibliography{Bibliography}

\bibliographystyle{unsrt}

\end{document}